\documentclass[3p]{elsarticle}

\usepackage{amssymb,amsmath}
\usepackage{bm}
\usepackage{graphicx}
\usepackage{tabularx}
\usepackage{comment}
\usepackage{fancyvrb}

\includecomment{pdffigure}

\DefineVerbatimEnvironment{code}{Verbatim}{fontsize=\small}
\newcommand{\openone}[0]{\leavevmode\hbox{\small1\normalsize\kern-.33em1}}

\biboptions{sort&compress}

\begin{document}

\begin{frontmatter}


\title{OCTBEC -- A Matlab toolbox for optimal quantum control of Bose-Einstein condensates}

\author{Ulrich Hohenester}
\ead{ulrich.hohenester@uni-graz.at}
\ead[url]{http://physik.uni-graz.at/~uxh}
\address{Institut f\"ur Physik. Karl--Franzens--Universit\"at Graz, 
  Universit\"atsplatz 5, 8010 Graz, Austria}



\begin{abstract}
\texttt{OCTBEC} is a Matlab toolbox designed for optimal quantum control, within the framework of optimal control theory (OCT), of Bose-Einstein condensates (BEC).  The systems we have in mind are ultracold atoms in confined geometries, where the dynamics takes place in one or two spatial dimensions, and the confinement potential can be controlled by some external parameters.  Typical experimental realizations are atom chips, where the currents running through the wires produce magnetic fields that allow to trap and manipulate nearby atoms.  The toolbox provides a variety of Matlab classes for simulations based on the Gross-Pitaevskii equation, the multi-configurational Hartree method for bosons, and on generic few-mode models, as well as optimization problems.  These classes can be easily combined, which has the advantage that one can adapt the simulation programs flexibly for various applications.
\end{abstract}


\date{September 22, 2013}

\end{frontmatter}

\section*{Program summary}

\noindent{\sl Program title:} \texttt{OCTBEC}\\
\noindent{\sl Programming language:} Matlab 7.11.0 (R2010b)\\
\noindent{\sl Computer:} Any which supports Matlab 7.11.0 (R2010b)\\
\noindent{\sl Operating system:} Any which supports Matlab 7.11.0 (R2010b)\\
\noindent{\sl RAM required to execute with typical data:} $\ge 1$ GByte\\
\noindent{\sl Has the code been vectorised or parallelized?:} no\\
\noindent{\sl Keywords:} Bose-Einstein condensates, optimal quantum control, atom chips\\
\noindent{\sl External routines/libraries used:} none\\
\noindent{\sl Nature of problem:} Simulation of Bose-Einstein condensates and optimal quantum control\\
\noindent{\sl Solution method:} Gross-Pitaevskii equation, multi-configurational Hartree method for bosons, generic few-mode models\\
\noindent{\sl Running time:} between seconds and hours\\

\section{Introduction}\label{sec:intro}

Bose-Einstein condensates and ultracold atoms in atom chips provide an ideal laboratory for the study of quantum physics under well-controlled conditions.  The possibility to store, manipulate \cite{guenther:05,luo:04,krueger:03,schumm:05,hofferberth:06,wang:05}, and measure single quantum systems with extremely high precision has initiated great stimulus in various fields of research, ranging from atom interferometry \cite{haensel:01,andersson:02,wang:05,schumm:05}, over quantum gates \cite{calarco:00,charron:06,treutlein:06} and resonant condensate transport \cite{paul:05}, to nonlinear atom optics \cite{deng:99,orzel:01,campbell:06,perrin:07}.  In the vast majority of these schemes the wavefunction of the Bose-Einstein condensate, trapped in the vicinity of an atom chip \cite{reichel:11}, is manipulated through variation of the magnetic confinement potential.  This is achieved by changing the currents through the gate wires mounted on the chip or modifying the strength of additional radio-frequency fields \cite{folman:02,lesanovsky:06,lesanovsky:06b,hofferberth:06,wildermuth:06}.  These external, time-dependent parameters thus provide a versatile control for wavefunction manipulations, and make atom chips attractive candidates for quantum control applications.

In the past, we have successfully developed and implemented an optimal quantum control approach for Bose-Einstein condensates in magnetic microtraps within the framework of the Gross-Pitaevskii equation \cite{hohenester.pra:07,borzi:08}.  More recently, this approach has been generalized to more sophisticated description schemes, such as the two-mode model \cite{milburn:97,javanainen:99} or the multi-configurational Hartree method for bosons (MCTDHB) \cite{alon:08}, which allow to additonally describe condensate fragmentations and excitations.  Atom number-squeezing and atom interferometry at the Heisenberg limit has been studied in a number of papers \cite{grond.pra:09,grond.pra:09b,hohenester.fdp:09,grond.njp:10,grond:11,grond:11b}.  Most recently, we have presented first results that show that optimal quantum control protocols can be successfully implemented in experiment \cite{buecker:11,buecker:12,buecker:13}.

Our computational approach is based on a Runge-Kutta or Crank-Nicolson solution of the dynamic equations, together with an optimal control framework using conjugate gradient or quasi-Newton minimization techniques~\cite{hohenester.pra:07,borzi:08,vonwinckel:08,grond.pra:09b,jaeger:13}.  Despite the large diversity of problems studied in the past, the simulations are similar in many aspects, and at some point we realized that we were wasting a significant amount of time copying code from one program to another one.  

The purpose of the \texttt{OCTBEC} toolbox is to provide a unified platform for the simulation of Bose-Einstein condensates and quasi-condensates in restricted geometries, where the confinement potential can be modified by some external \textit{control parameters}, and to allow for the optimization of the time variation of these control parameters within the framework of optimal control theory.  Typical simulation scenarios consist of a few hundreds to thousands of atoms, with confinement lengths and manipulation times of the order of micrometers and milliseconds, respectively.  A modular structure of the program has been achieved by using \texttt{classdef} objects available with Matlab 7.6 or higher versions.  These classes can be easily combined such that one can adapt the simulation programs flexibly to the user's needs.  A comprehensive help is available for all classes and functions of the toolbox through the \texttt{doc} command.  In addition, we have created detailed help pages, accessible in the Matlab help browser, together with a complete list of the classes and functions of the toolbox, and a number of demo programs.  In this paper we provide an ample overview of the \texttt{OCTBEC} toolbox, but leave several details to the help pages.  

The model systems covered by the toolbox include:

\begin{description}

\item[] \textit{Gross-Pitaevskii.}
The Gross-Pitaevskii equation describes the condensate dynamics in terms of a single wavefunction, and the non-linear atom-atom interactions are accounted for through a mean-field approach~\cite{dalfovo:99,leggett:01}.  The Gross-Pitaevskii framework proves to be extremely successful for problems where condensate fragmentation or excitation are of minor importance.

\item[] \textit{Few-mode model.}
In the few-mode model, atoms become distributed between static or time-dependent orbitals and the time evolution is governed by a Hamiltonian matrix \cite{milburn:97,javanainen:99}.  The whole condensate dynamics is then associated with the atom number distribution, whereas the orbital degrees of freedom are lumped into a few effective parameters.  

\item[] \textit{MCTDHB.}
In the multiconfigurational time-dependent Hartree method for bosons (MCTDHB) one accounts for both the spatial and atom-number dynamics.  The approach has been developed by Cederbaum, Alon, and coworkers \cite{alon:08}, and allows for an ab-initio description of the condensate dynamics, at least in principle.  We refer the interested reader to \texttt{www.pci.uni-heidelberg.de/tc/usr/mctdhb/} where a collection of papers and software programs can be found.

\end{description}

In the toolbox we provide implementations for the simulation of the above models.  In comparison to the \texttt{OpenMCTDHB} software provided by the Cederbaum group, our implementation is probably less refined, but can be used in combiation with optimal control applications.  As regarding the structure of the toolbox and the philosophy behind its implementation, a number of comments are at place.

Whenever possible, we have tried to favor readability and transparancy over runtime and memory requirements.  It is likely that the performance of most programs could be significantly improved, yet we have tried to stick to a modular structure and to strict programming rules throughout.  We hope that with this approach the programs are easier to read, and that other model systems can be implemented without too much knowledge about the working principles of the remaining toolbox.  Quite generally, our primary interest is to apply the software to physically interesting problems, rather than to develop software and methodology.  In this respect, a flexible software platform is extremely helpful because it allows to build on already established expertise, and to devote work and time to the novel aspects of a given problem only. 

This paper is organized as follows.  In Sec.~\ref{sec:overview} we briefly review the different model systems and provide a short overview of the toolbox.  The toolbox installation and a few selected examples are discussed in Sec.~\ref{sec:start}.  In Sec.~\ref{sec:preliminaries} we describe the unit system, the computational grid, the control parameters, and the ODE solvers provided by the toolbox.  The implementation of the different model systems is explained in Sec.~\ref{sec:model}.  Finally, optimal control theory and its implementation are explained in Sec.~\ref{sec:oct} and applied to the different model systems in Sec.~\ref{sec:optmodel}.  Details about some of our Crank-Nicolson and OCT implementations can be found in the Appendices.

\section{Theory and brief overview}\label{sec:overview}

\subsection{Model systems}

Our starting point for the description of condensate dynamics in restricted geometries is the many-body Hamiltonian in second-quantized form~\cite{dalfovo:99,leggett:01}, which, for simplicity, we give for a one-dimensional system,
\begin{equation}\label{eq:ham2quant}
  \hat H=\int\hat\Psi^\dagger(x)\left[
  -\frac{\hbar^2}{2M} \frac{\partial^2}{\partial x^2}+V(x,\lambda(t))\right]\hat\Psi(x)\,dx+
  \frac{\kappa}2\int \hat\Psi^\dagger(x)\hat\Psi^\dagger(x)\hat\Psi(x)\hat\Psi(x)\,dx\,.
\end{equation}
The first term on the right-hand side accounts for the kinetic energy, where $\hbar$ is the reduced Planck constant and $M$ the atom mass, as well as for the magnetic confinement potential $V(x,\lambda(t))$.  The control parameter $\lambda(t)$ determines the variation of the confining potential when changing the external parameters \cite{folman:02,lesanovsky:06} (for details see below).  Through $\lambda(t)$ it is possible to manipulate the trapped Bose-Einstein condensate, e.g. to split it by varying the potential from a single to a double well, or to excite it by displacing the potential minimum.  The second term on the right-hand side accounts for the atom-atom interactions, where we have chosen a contact potential approximation for the interatomic potential \cite{dalfovo:99,leggett:01}.  The bosonic field operators $\hat\Psi(x)$ and $\hat\Psi^\dagger(x)$ obey the usual equal-time commutation relations.  

Different physical regimes emerge from Eq.~\eqref{eq:ham2quant} by restricting the field operators to certain types of basis functions.  First, the \textit{Gross-Pitaevskii equation} \cite{dalfovo:99} is obtained by assuming that all atoms reside in a single ``orbital'' $\psi(x,t)$. Correspondingly, the field operator is
\begin{equation}
  \hat\Psi(x)=\hat a_0\,\psi(x,t)\,,
\end{equation} 
where $\hat a_0$ is the operator associated with the condensate. The Gross-Pitaevskii equation properly accounts for the mean-field dynamics of the condensate, described by the orbital $\psi(x,t)$, but cannot cope with correlation effects and fragmentation, where more than a single orbital becomes populated.

A prominant example is splitting of a condensate, which can be achieved by transforming the confinement potential from a single to a double well, where at some point the condensate breaks up into two parts, $\psi_L(x)$ and $\psi_R(x)$, which are localized in either the left or right well.  Close to the splitting point, we can approximately ignore the dynamics of the orbitals $\psi_{L,R}(x)$, and recover the \textit{two-mode model} \cite{milburn:97,javanainen:99}
\begin{equation}\label{eq:field.twomode}
  \hat\Psi(x)=\hat a_L\,\psi_L(x)+\hat a_R\,\psi_R(x)\,.
\end{equation} 
Here, the whole condensate dynamics is associated with the atom number distribution, through the field operators $\hat a_L$ and $\hat a_R$, whereas the orbital degrees of freedom are lumped into a few effective parameters.  We will refer to such kind of systems as \textit{few-mode models}.  Quite generally, such models can also describe the atom dynamics in optical lattices \cite{morsch:06} within the Bose-Hubbard framework \cite{jaksch:98}.

In the most general case, we can neither neglect the orbital nor the atom number dynamics. This can be done by choosing a field operator of the form
\begin{equation}\label{eq:field.mctdhb}
  \hat\Psi(x)=\hat a_L(t)\,\psi_L(x,t)+\hat a_R(t)\,\psi_R(x,t)\,,
\end{equation} 
where $\psi_{L,R}(x,t)$ are time-dependent orbitals that have to be determined self-consistently.  A convenient approach is provided by the \textit{multi-configurational time dependent Hartree method for bosons (MCTDHB)} \cite{alon:08} which determines the orbitals from a variational principle.

\begin{figure}[t]
\begin{pdffigure}
\centerline{\includegraphics[width=0.8\columnwidth]{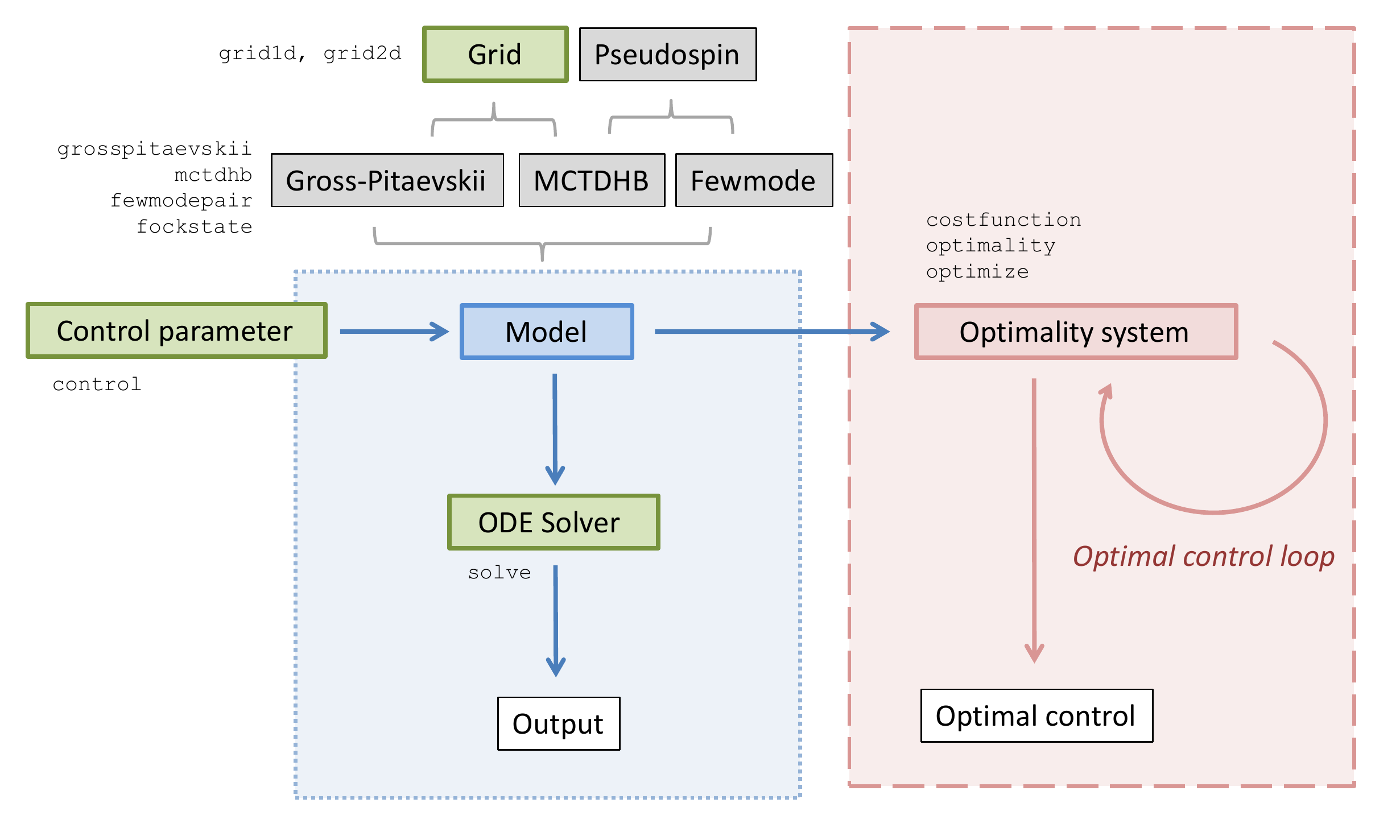}}
\end{pdffigure}
\caption{Overview of the OCTBEC Toolbox for the solution of the Gross-Pitaevskii, MCTDHB, and fewmode models.  An external control parameter allows to modify the confinement potential or other model parameters through which the wavefunction can be steered.  Within the toolbox one can either solve the dynamic equations, or submit the problem to optimal control theory.  Here an \textit{optimal control} is determined, which minimizes a cost function that parameterizes the control objective.  In the figure we add class and function names that can be assigned to the specific tasks.
}\label{fig:flowchart}
\end{figure}

\subsection{Brief overview}

The main purpose of the OCTBEC toolbox is to provide a flexible toolkit for the simulation of condensate dynamics in confined geometries, where the confinement can be modified by some external control parameters.  The theoretical frameworks covered by the toolbox go from the Gross-Pitaevskii model, over fewmode models, to the multi-configurational Hartree method for bosons.  Figure~\ref{fig:flowchart} provides a detailed overview of the different tasks which can be separated into (i) simulation of the dynamics, and (ii) quantum control within the framework of optimal control theory.  In the following we discuss prototypical examples and provide details about the different classes and functions of the toolbox.

\section{Getting started}\label{sec:start}

\subsection{Installation of the toolbox}

\begin{figure}[t]
\begin{pdffigure}
\centerline{\includegraphics[width=0.9\columnwidth]{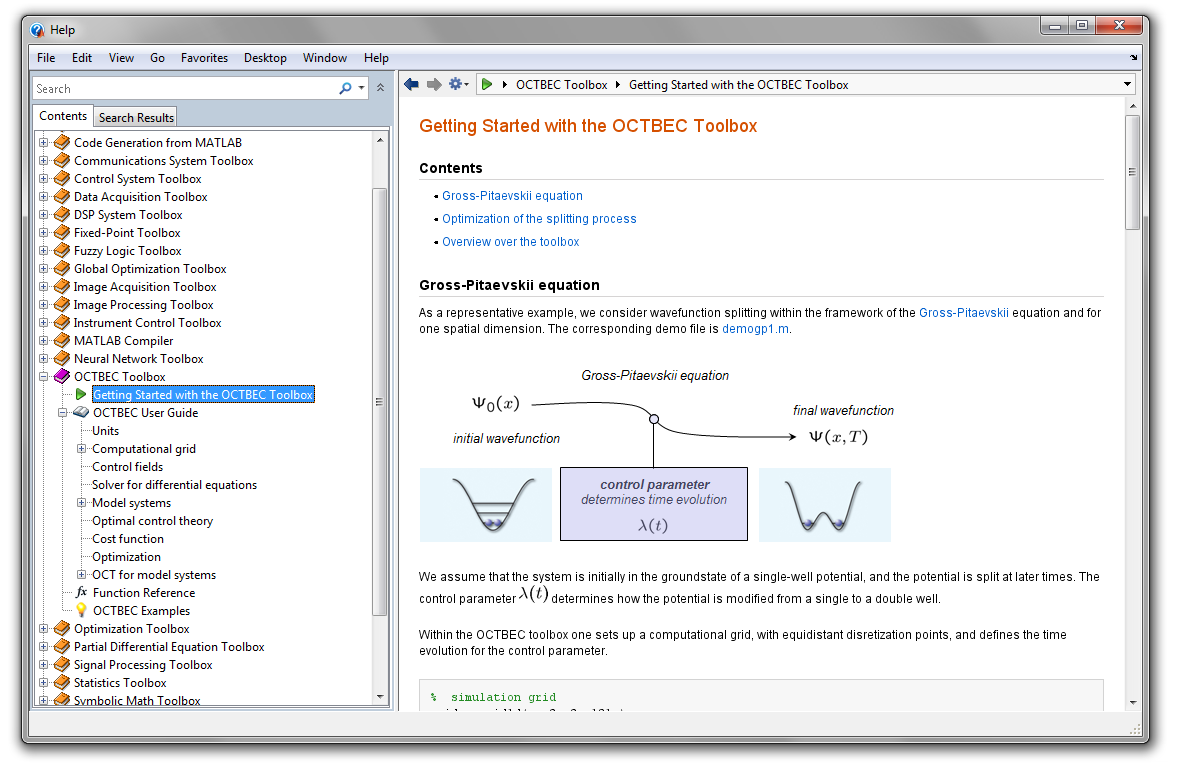}}
\end{pdffigure}
\caption{Screenshot of the help pages of the \texttt{OCTBEC} toolbox within the Matlab help browser.  The help pages provide a short introduction, a detailed user guide, a list of the classes and functions of the toolbox, as well as a number of demo programs.}\label{fig:helppages}
\end{figure}

To install the toolbox, one must simply add the path of the main directory \texttt{octbecdir} of the \texttt{OCTBEC} toolbox as well as the paths of all subdirectories to the Matlab search path.  This can be done, for instance, through
\begin{code}
addpath(genpath(octbecdir));
\end{code}
To set up the help pages, one must once change to the main directory of the \texttt{OCTBEC} toolbox and run the program \texttt{makeoctbechelp}

\begin{code}
>> cd octbecdir;
>> makeoctbechelp;
\end{code}
Once this is done, the help pages, which provide detailed information about the toolbox, are available in the Matlab help browser.  Note that one may have to call \texttt{Start > Desktop Tools > View Start Button Configuration > Refresh} to make the help accessible.  Under Matlab 2012 the help pages can be found on the start page of the help browser under \textit{Supplemental Software}.  Figure~\ref{fig:helppages} shows a screenshot of the \texttt{OCTBEC} help pages.  This manuscript closely follows the help pages, but additionally provides further details about the theory and methodology underlying our computational approach.

\subsection{A few selected examples}\label{sec:examples}

\begin{figure}
\centerline{\includegraphics[width=0.3\columnwidth]{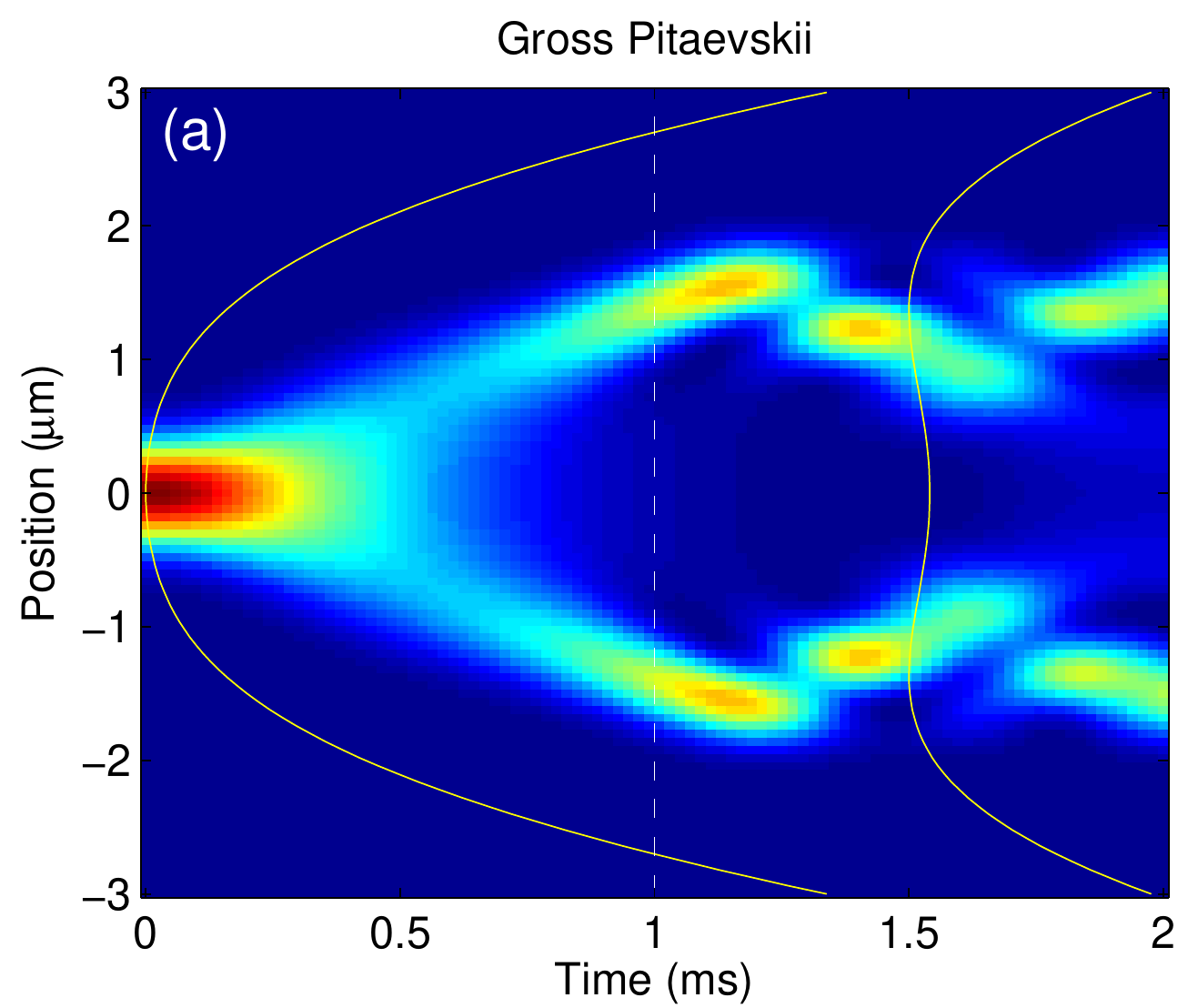}\quad
            \includegraphics[width=0.3\columnwidth]{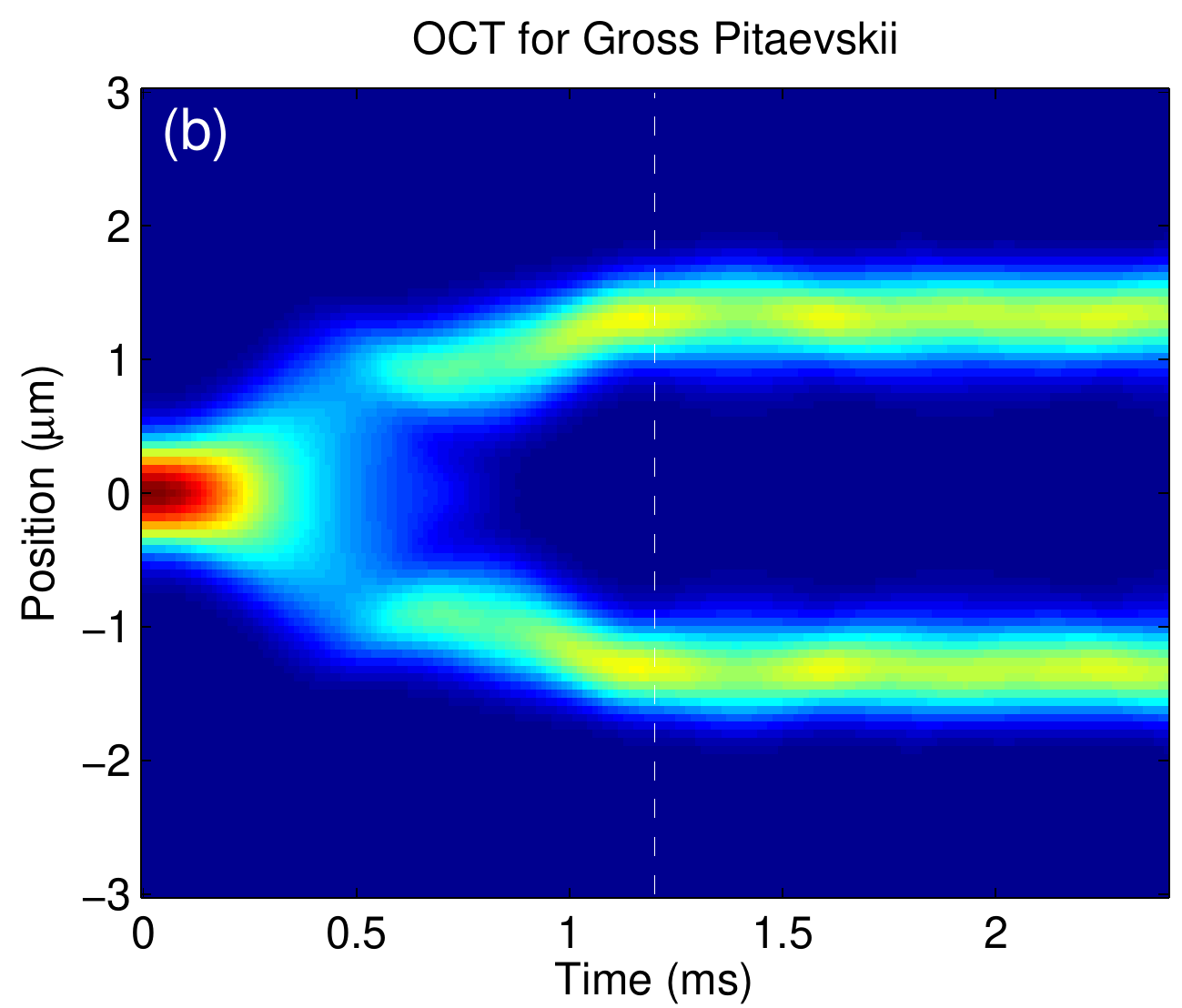}\quad
            \includegraphics[width=0.3\columnwidth]{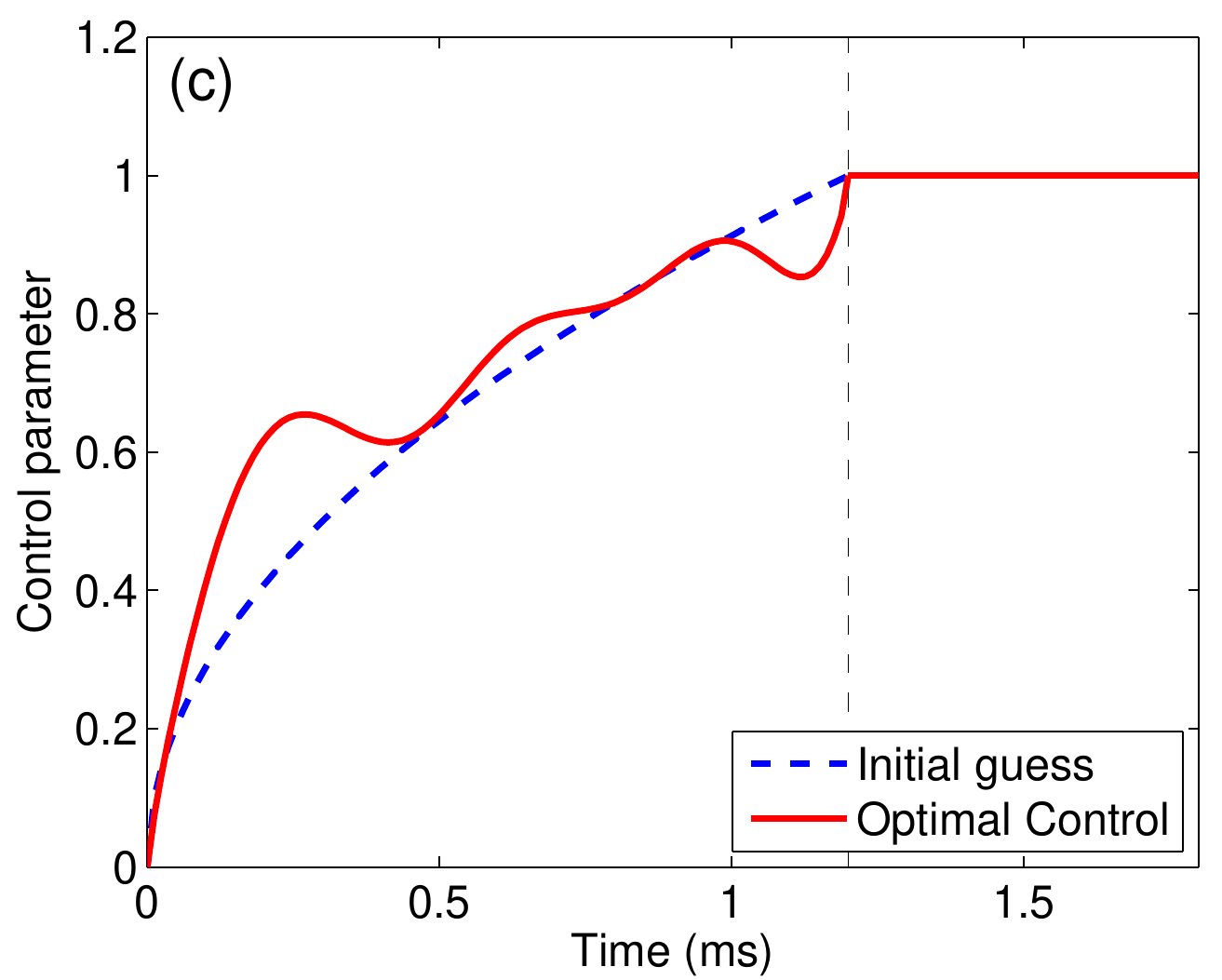}}
\caption{Results of the demo programs (a) \texttt{demogp1.m} and (b,c) \texttt{demogpoct1.m}.  (a) Particle density for a splitting process where the confinement potential (yellow lines) is transformed from a single to a double well.  After 1 ms the confinement potential is kept constant and the condensate oscillates within the separated minima of the double well potential.  (b) Particle density evolution for an optimized control parameter, which brings the condensate to the groundstate at the terminal time $T=1.2$ ms of the splitting process.  (c) Initial guess for time evolution of control parameter (dashed line) and optimized $\lambda(t)$.}\label{fig:examples}
\end{figure}

To get a first idea what the toolbox can do, it is probably best to work through some of the demo programs listed in the \textit{OCTBEC Examples} section of the help pages.  By typing 
\begin{code}
>> demogp1
\end{code}
at the Matlab prompt, a simulation is performed where a condensate wavefunction is split by transforming the potential from a single to a double well (the demo file can be opened in the Matlab editor with \texttt{edit demogp1.m}).  Results are shown in Fig.~\ref{fig:examples}(a).  Such splitting has been described in some length in Ref.~\cite{hohenester.pra:07}, and details of the demo program will be presented in Sec.~\ref{sec:gp}.  

The demo program \verb|demomctdhb1| investigates a similar splitting scenario, however within the framework of the MCTDHB(2) equations~\cite{alon:08}.  The program performs somewhat slower, which is due to the complexity of the underlying equations that account for both the orbital dynamics and the distribution of atoms between these orbitals.  For the investigated splitting scenario the results are very similar to those of the Gross-Pitaevskii equation.

Finally, we briefly discuss an optimal control simulation where the protocol for the splitting process is optimized such that the condensate ends up in the groundstate of the split trap at the terminal time.  Such splitting has been investigated in Ref.~\cite{hohenester.pra:07}, and details will be presented in Sec.~\ref{sec:optgp}.  We start the program with
\begin{code}
>> demogpoct1    %  runtime about 90 sec
\end{code}
The program opens a graphics window that shows the iterative improvement of the control parameter.  In addition, in the Matlab window the progress of the optimal control loop is reported
\begin{code}
it=  1   f=1.518970e-001   ||g||=9.910656e-001   sig=0.125
it=  2   f=5.713967e-002   ||g||=1.527621e-001   sig=0.140
it=  3   f=4.720535e-002   ||g||=9.593845e-003   sig=0.126
it=  4   f=2.144892e-002   ||g||=3.831657e-002   sig=5.144
it=  5   f=1.829915e-002   ||g||=3.222441e-003   sig=0.951
it=  6   f=1.782431e-002   ||g||=5.266112e-004   sig=1.363
it=  7   f=1.699185e-002   ||g||=4.061750e-003   sig=5.821
it=  8   f=1.456352e-002   ||g||=5.018247e-004   sig=3.480
it=  9   f=1.451606e-002   ||g||=3.001156e-004   sig=0.522
\end{code}
Here, \verb!it! gives the iteration number of the optimization loop, \verb!f! is the \textit{cost function} value~\cite{hohenester.pra:07}, and \verb!||g||! is the norm of the gradient which should become zero for the optimal control.  Figs.~\ref{fig:examples}(b,c) show the time evolution of the particle density and the initial and optimal control, respectively.  As can be seen, for the optimal control the condensate ends up close to the groundstate of the split trap, and only minor oscillations occur at later times when the trap is kept constant.

\begin{table}[t]
\caption{Selected examples provided by the OCTBEC toolbox.  We list the names of the programs, typical runtimes, and give brief explanations.  The programs were tested on a standard PC (Intel i7--2600 CPU, 3.40 GHz, 8 GB RAM).  The different runtimes for Gross-Pitaevskii and MCTDH simulations can be infered from the comparison of \texttt{demogp1} and \texttt{demomctdhb1}, as well as \texttt{demogpoct1} and \texttt{demomctdhboct1} for OCT simulations.}\label{table:examples}
{\small
\begin{tabularx}{\columnwidth}{lrX}
\hline\hline
Demo program & Runtime & Description \\
\hline
\texttt{demogp1} & 0.75 sec & Splitting of condensate wavefunction within Gross-Pitaevskii framework\\
\texttt{demogp1split} & 0.57 sec & Same as \texttt{demogp1} but using split-operator integration\\
\texttt{demogp2} & 0.80 sec & Solution of 2d Gross-Pitaevskii equation with split operator \\
\texttt{demogp3} & 4.47 sec & Shake-up process within Gross-Pitaevskii framework \\
\texttt{demogpoct1} & 89.02 sec & OCT simulation of condensate splitting within Gross-Pitaevskii framework\\
\texttt{demogpoct2} & 93.21 sec & Same as \texttt{demogpoct1} but energy minimization for cost function\\
\texttt{demogpoct3} & 350.06 sec & OCT of shake-up process within Gross-Pitaevskii framework \\
\texttt{demomctdhb1} & 4.91 sec & Splitting of condensate wavefunction within MCTDHB framework\\
\texttt{demomctdhboct1} & 942.27 sec & Same as \texttt{demogpoct1} but for MCTDHB model\\
\texttt{demomctdhboct2} & 853.09 sec & Same as \texttt{demogpoct2} but for MCTDHB model\\
\texttt{demofewmodepair} & 1.42 sec & Number squeezing through splitting\\
\texttt{demofewmodepairbloch} & 11.31 sec & Visualization of number squeezing on Bloch sphere\\
\hline
\hline
\end{tabularx}}
\end{table}

\begin{table}
\caption{A few selected properties and methods of the \texttt{grid1d} class. Use \texttt{doc @grid1d} to get a complete listing of all class properties and methods.}
\label{table:grid1d}
{\small
\begin{tabularx}{\columnwidth}{ccX}
\hline\hline
Property & Type & Description \\
\hline
\verb|n| & Integer      & Number of positions\\
\verb|x| & Double array & Positions of grid\\
\verb|grad|  & Sparse matrix & Derivative operator\\
\verb|grad4| & Sparse matrix & Derivative operator (4th order accuracy)\\
\verb|lap|   & Sparse matrix & Laplace operator\\
\verb|lap4|  & Sparse matrix & Laplace operator (4th order accuracy)\\
& & \\
\verb|norm| & Function & Norm of wavefunction \\
\verb|normalize| & Function & Normalize wavefunction \\
\verb|inner| & Function & Inner product of two wavefunctions \\
\verb|integrate| & Function & Integrate function on grid \\
\hline
\hline
\end{tabularx}}
\end{table}

Table \ref{table:examples} lists a number of additional demo programs, and provides typical runtimes as well as a brief explanation.  Deatils about the classes and functions can be found in the help pages of the toolbox or by typing \verb!doc @classname! at the Matlab prompt.

\section{Preliminaries}\label{sec:preliminaries}

In this section we introduce several basic classes and concepts needed for the simulation of BECs in confined geometries.  We first briefly describe the unit system adopted in the demo files (the toolbox itself does not rely on a specific unit system), and then introduce the computational grid used for the discretization of the spatial domain and our implementation of control parameters.  Finally, we present the solvers for ordinary differential equations provided by the toolbox.

\subsection{Units}\label{sec:units}

The toolbox does not use a specific unit system, and the choice of units is in principle left to the user.  In the demo files we use units where length is measured in micrometers, time is measured in milliseconds, and the reduced Planck constant is set to $\hbar=1$.  With this choice, atom masses have to be given in units of 
\begin{equation}
  M_{\rm nucl} / (\hbar\times L)\,,
\end{equation}
with $M_{\rm nucl}$ being the nucleon mass and $L=1\,\mu m$ the length unit.  For instance, to properly set the mass of Rubidium atoms one then proceeds as follows:

\begin{code}
units;                %  load units (defines mass)
massRb = 87 * mass;   %  mass of Rubidium atoms (mass number 87)
\end{code}
We believe that this unit system is well suited for the problems under study, and recommend to use it whenever possible.

\subsection{Spatial grid}\label{sec:grid}

\subsubsection{One-dimensional grid}

In our computational approach, we represent the spatial domain by a grid of discrete points and approximate function derivatives by finite differences.  Consider the situation where the one-dimensional domain $x\in [x_{\rm min},x_{\rm max}]$ is represented by $n$ discrete points.  Within the OCTBEC toolbox, one calls
\begin{code}
grid = grid1d( xmin, xmax, n );   %  initialization of computational grid in 1d
\end{code}
Upon initialization of \verb|grid|, a Matlab object is created whose properties are listed in table~\ref{table:grid1d} (use \verb|doc @grid1d| to get a complete listing of all class properties and methods).  The derivative operators, such as \verb|grad| or \verb|lap|, can be applied to functions through simple multiplication.  Let us consider the example of Rubidium atoms inside a harmonic trap.  To compute the condensate groundstate in absence of nonlinear atom-atom interactions, we proceed as follows.

\begin{figure}
\centerline{\includegraphics[width=0.5\columnwidth]{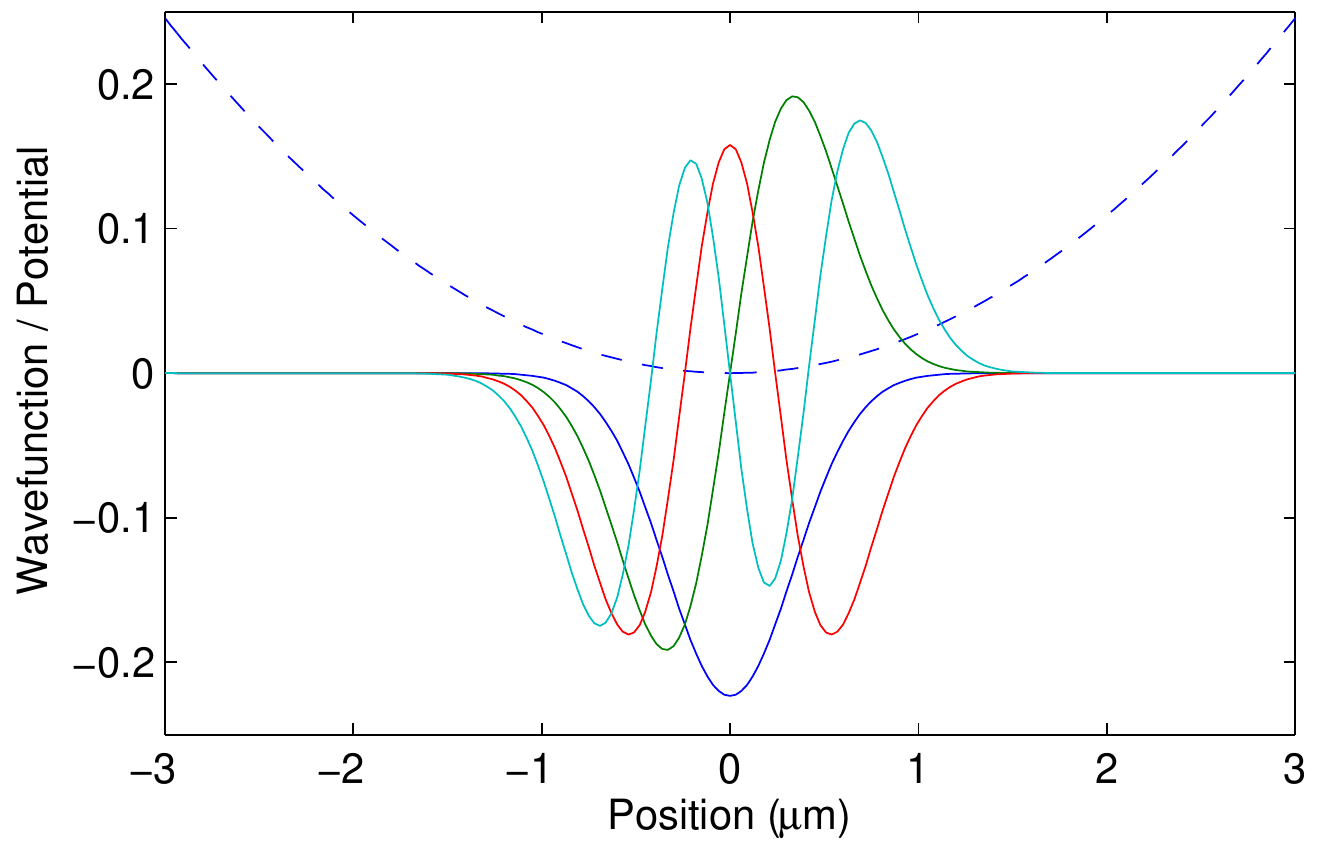}}
\caption{Confinement potential (blue dashed line) and eigenfunctions (solid lines) for a harmonic confinement potential.}\label{fig:grid1d}
\end{figure}

\begin{code}
units;                          %  length in micrometers, time in ms, and hbar = 1
grid = grid1d( - 3, 3, 201 );   %  initialize grid

%  kinetic energy
t = - 0.5 * grid.lap4 / ( 87 * mass );
%  potential energy for 1 kHz confinement potential
v = 0.5 * ( 87 * mass ) * ( 2 * pi ) ^ 2 * spdiag( grid.x ) .^ 2;
%  compute the four eigenfunctions and eigenvalues of lowest energy
[ psi, ene ] = eigs( t + v, 4, 'sa', struct( 'disp', 0 ) );

%  plot eigenfunctions and confinement potential
plot( grid.x, 1e-3 * diag( v ), 'b--', grid.x, psi );
\end{code}
We have used the toolbox function \verb|spdiag(a)| to place the values of a vector \verb|a| on the diagonal of a sparse matrix.  Figure~\ref{fig:grid1d} shows the confinement potential and eigenfunctions for this harmonic confinement potential.  The \verb|grid1d| class provides a number of methods that can be used to manipulate wavefunctions.  Below we show several examples.

\begin{code}
%  norm of wavfunction 
grid.norm( psi )                        %  0.1732    0.1732    0.1732    0.1732                   
%  normalize wavefunctions  
psi = grid.normalize( psi );
%  norm of wavefunctions
grid.integrate( abs( psi ) .^ 2 )       %  1.0000    1.0000    1.0000    1.0000
%  inner product of two wavefunctions 
grid.inner( psi( :, 1 ), psi( :, 2 ) )  %  -4.2279e-017  
\end{code}

\subsubsection{Two-dimensional grid}

The toolbox provides with \verb!grid2d! also a two-dimensional grid.  Similarly to the one-dimensional case, initialization is done via
\begin{code}
grid = grid2d( xmin, xmax, nx, ymin, ymax, ny );  %  initialization of 2d mesh
\end{code}
Most methods and functions are similar to \verb!grid1d!, in addition the structure \verb!grid.mesh! provides the $x$ and $y$ coordinates obtained from \verb!meshgrid! for plotting within Matlab (e.g. through \verb!imagesc!).  A more detailed explanation can be found in the help pages, or can be obtained through \verb!doc @grid2d!.  At present we do not provide a class for grids in all three spatial dimensions.

\subsection{Control parameter}\label{sec:control}

The \texttt{OCTBEC} toolbox considers situations where the confinement potential for atoms can be controlled by some external \textit{control parameters} $\lambda(t)$ \cite{hohenester.pra:07}.  A typical example are the magnetic fields of an atom chip, where the potential can be modified by changing the currents running through the wires of the chip.  To properly access the classes and functions of the toolbox, one usually has to specify the time variation of the control field (or the control fields, if needed), which we assume to be real-valued throughout.  To this end, one calls:

\begin{code}
ttab = linspace( 0, 3, 100 );        %  control field tabulated at given times
lamtab = 0.1 * ttab;                 %  time variation of control field
lambda = control( ttab, lamtab );    %  set up control field

lambda( 0.5 )       %  0.05, interpolate control field at intermediate time
\end{code}
Here we use tabulated values to set up the control object.  If several control fields are needed, \verb|lamtab| should be a matrix rather than an array.  The control object can perform interpolations between the tabulated values.  Note that \verb|lam=lambda(t)| returns a \verb|controltype| object which behaves very much like a double variable or double array, depending on the dimension of the control parameter.  However, through \verb|lam.t| one can additionally access the time at which the control parameter is evaluated, which might be useful for problems with an explicit time dependence.  The full strength of the \verb|control| objects will become clear in connection with optimal control theory (OCT), which seeks for optimal variations of the control fields in order to fulfill certain control objectives.

\begin{figure}
\centerline{\includegraphics[width=0.5\columnwidth]{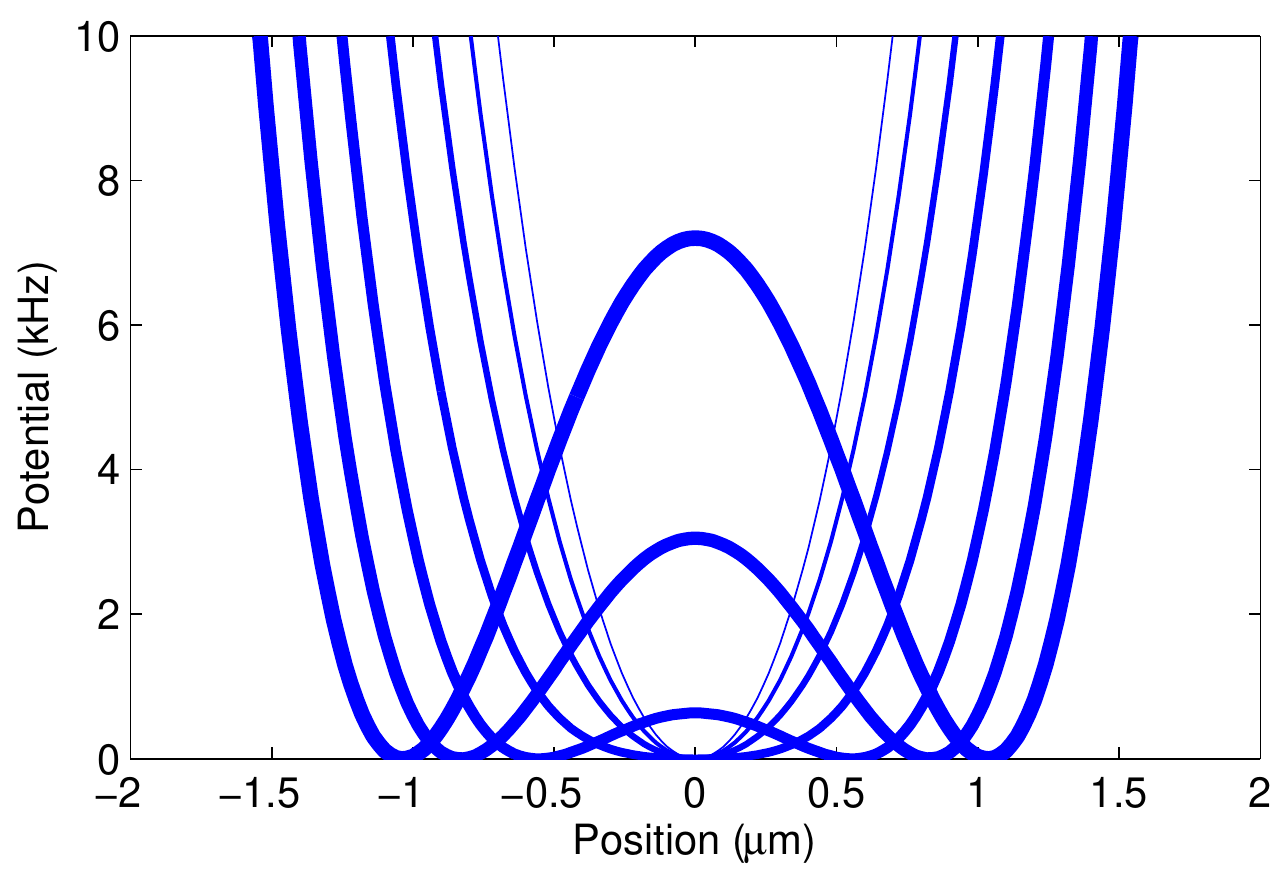}}
\caption{Confinement potentials for different values of the control parameter $\lambda(t)$ and for the potential considered by Lesanovsky et al. \cite{lesanovsky:06} which describes the transition from a single to a double well.}\label{fig:control}
\end{figure}

Usually the control parameter works together with a function or function handle for the confinement potential.  For instance,

\begin{code}
v = @( lambda ) ( 0.5 * k * ( grid.x - lambda ) .^ 2 );  %  harmonic confinement potential
\end{code}
defines a harmonic confinement potential whose origin is shifted by the control parameter.  As another example, we consider the potential of Lesanovsky et al. \cite{lesanovsky:06} for the transition from a single well to a double well, as shown in Fig.~\ref{fig:control}.

\begin{code}
grid = grid1d( - 3, 3, 301 );                        %  set up computational grid
v = @( lambda ) ( lesanovsky1d( grid.x, lambda ) );  %  confinement potential

tout = linspace( 0, 4, 201 );                              %  simulation times
lambda = control( tout, 0.2 + tout / max( tout ) * 0.6 );  %  temporal variation control parameter

%  loop over times
for t = linspace( 0, max( tout ), 7 )
  plot( grid.x, v( lambda( t ) ), 'LineWidth', 0.5 + t );  hold on;
end
\end{code}

\subsection{ODE solver}\label{sec:odesolver}

The OCTBEC toolbox provides a solver for ordinary differential equations (ODEs) that works together with generic Matlab classes.  To work properly, the following operations and functions must be implemented for the class:

\medskip
\begin{tabular}{ll}
\texttt{plus} & add two objects together, e.g. \verb|obj1+obj2|, \\
\texttt{mtimes} & multiply object with constant value, e.g. \verb|obj*val|,\\
\texttt{deriv} & compute derivative function for ODE,\\
\texttt{crank} & perform Crank-Nicolson step.\\
\end{tabular}
\medskip

\noindent The function \verb|deriv| is used by the Runge-Kutta solver and the function \verb|crank| by the Crank-Nicolson solver.  The calls to these functions are of the form
\begin{code}
dy = deriv( y, lambda, ~ );           %  derivative function for Runge-Kutta solver
ynew = crank( yold, lambda, dt, ~ );  %  Crank-Nicolson step
\end{code}
Here \verb|y| and \verb|yold| are user-defined objects, \verb|lambda| is a control parameter, and \verb|dt| is the step size of the Crank-Nicolson scheme.  Note that the time argument can be accessed via \verb|lambda.t|.  The last argument in the above functions can be used to pass options to the derivative functions.  The OCTBEC toolbox provides a number of classes that have already defined all of the above operations and functions.

Once the object is defined, one can solve the ODE through
\begin{code}
[ yout, tout ] = solve( y0, tout, lambda, op );  %  solve differential equation
\end{code}
Here \verb|y0| is the initial value, \verb|tout| are the times where the output is requested, \verb|lambda| is a control object, and \verb|op| is an option structure that controls the ODE integration.  The ODE solver returns a cell array \verb|yout| of output values.  The option structure can have the following fields

\medskip
\begin{tabular}{ll}
\texttt{nout} & intermediate output (waitbar) after \texttt{nout} timesteps, \\
\texttt{nsub} & subdivide each timestep into \texttt{nsub} sub-timesteps.\\
\texttt{stepfun} & \verb|'crank'| for Crank-Nicolson, \verb|'runge4'|, or any Matlab ODE solver such as \verb|'ode23',|\\
\texttt{funiter} & function to be called after each time step,\\
\texttt{funout} & function to be called at output steps.\\
\end{tabular}
\medskip

\noindent Through \verb|nsub| it is possible to refine the numerical integration without modifying the output results.  Note that with the \verb|'runge4'| option a 4th-order Runge-Kutta integration with fixed time steps will be performed.  The setting of \verb!nsub! is decisive only for the ODE solvers \verb!crank! and \verb!runge4! with fixed stepsize, for the builtin Matlab ODE solvers \verb!nsub! simply determines the initial stepsize.  In general, we recommend to either use the \verb!'crank'! option for fast and efficient integration with low accuracy, or \verb!'ode23'! which performs a sufficiently fast Runge-Kutta integration with adaptive step size.  The user-defined functions must be of the form
\begin{code}
y = funiter( t, y );    %  function to be called after each iteration
yout = funout( t, y );  %  function to be called for output
\end{code}
The function \verb|funiter| allows, for instance, to normalize a wavefunction after each time step, to avoid numerical rounding errors or to be used for imaginary time propagation, whereas the function \verb|funout| allows to save only part of the dynamic variables in the output.

\section{Model systems}\label{sec:model}

\begin{table}
\caption{Methods and properties for the different model classes.  Detailed help can be obtained by typing \texttt{doc @grosspitaevskii}, \texttt{doc @mctdhb}, or \texttt{doc @fockstate}.  For writing a new class, one has to implement the methods and properties listed under ``All''.
}\label{table:model}
{\small
\begin{tabularx}{\columnwidth}{llX}
\hline\hline
Model & Methods & Description \\
\hline
All & \texttt{plus, mtimes, mrdivide} & operations \texttt{+,*,/} \\
& \texttt{display} & Command window display of class properties \\
& \texttt{subsref} & Access class properties and functions\\
& \texttt{deriv} & Derivative function for Runge-Kutta solvers \\
& \texttt{crank} & Crank-Nicolson step \\
& \texttt{optderiv} & Derivative for OCT optimality system and Runge-Kutta integration \\
& \texttt{optcrank} & Crank-Nicolson step for OCT optimality system \\
& \texttt{derivpotential} & Derivative of Lagrange function with respect to control parameter \\
& \texttt{pack} & Pack wavefunction to column vector for use with Matlab ODE solvers \\
& \texttt{unpack} & Unpack wavefunction from column vector for use with Matlab ODE solvers \\
& & \\
Gross-Pitaevskii & \texttt{grid} & Real-space grid \\
& \texttt{ham} & Single-particle Hamiltonian \\
& \texttt{kappa} & Nonlinearity parameter \\
& \texttt{orbital} & Wavefunction of Gross-Pitaevskii object \\
& \texttt{density} & Particle density in real space\\
& \texttt{groundstate} & Gross-Pitaevskii groundstate \\
& \texttt{split} & Split operator step\\
& \texttt{optsplit} & Split operator step for optimality system\\
& & \\
MCTDHB & & All methods and properties of the Gross-Pitaevskii class are also implemented for the MCTDHB class, with the exception of \texttt{split} and \texttt{optsplit} \\
& \texttt{m} & Number of orbitals \\
& \texttt{n} & Number of atoms \\
& \texttt{num} & Atom number part of wavefunction\\
& \texttt{spin} & Pseudospin object for atom number part of wavefunction\\
& \texttt{densitymatrix} & One- and two-particle density matrices \\
& & \\
Fockstate & \texttt{num} & Atom number part of wavefunction\\
& \texttt{spin} & Pseudospin object for atom number part of wavefunction\\
& \texttt{ham} & Many-particle Hamiltonian \\
& \texttt{groundstate} & Groundstate of atom number wavefunction \\
\hline
\hline
\end{tabularx}}
\end{table}

Within the OCTBEC toolbox three model systems have been already implemented, namely the Gross-Pitaevskii, MCTDHB, and fewmode model.  All models have a similar structure but their own specialities, and can be submitted to the ODE solvers for solutions with either the Runge-Kutta or Crank-Nicolson stepper scheme.  In the following we provide details and examples for the different models.  Table~\ref{table:model} lists their methods and properties.

\subsection{Gross-Pitaevskii equation}\label{sec:gp}

\begin{figure}[t]
\begin{pdffigure}
\centerline{\includegraphics[width=0.5\columnwidth]{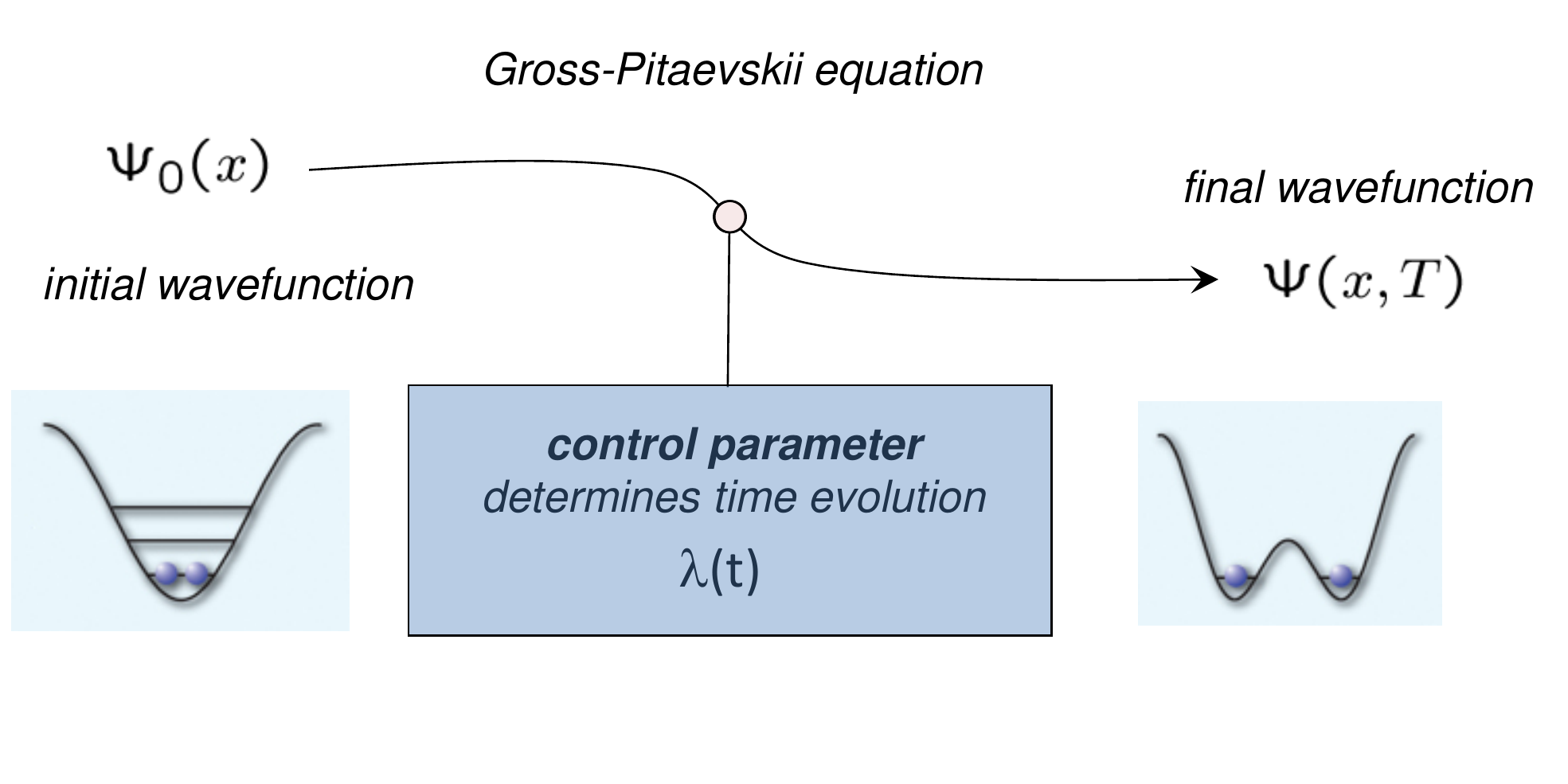}}
\end{pdffigure}
\caption{Schematics for solution of the Gross-Pitaevskii equation.  The confinement potential is modified from a single to a double well, where the details of the transition are controlled by the control parameter $\lambda(t)$.  $\psi_0(x)$ denotes the initial wavefunction and $\psi(x,T)$ the one at the end of the splitting process at time $T$.}\label{fig:gpschematics}
\end{figure}

The \verb|grosspitaevskii| object allows for the solution of the Gross-Pitaevskii equation~\cite{dalfovo:99,leggett:01}
\begin{equation}\label{eq:gp}
 i\hbar\frac{\partial\psi(\bm r,t)}{\partial t}=
   \Bigl(-\frac{\hbar^2\nabla^2}{2M} + V(\bm r,\lambda(t)) + 
   \kappa\bigl|\psi(\bm r,t)\bigr|^2\Bigr)\psi(\bm r,t)\,.
\end{equation}
The first term on the right-hand side is the operator for the kinetic energy, the second one is the confinement potential that can be controlled by some external parameter $\lambda(t)$, and the last term is the nonlinear atom-atom interaction in the mean field approximation of the Gross-Pitaevskii framework.  $M$ is the atom mass and $\kappa$ is the strength of the atom-atom interactions.  To set up the \verb|grosspitaevskii| object one calls

\begin{code}
psi = grosspitaevskii( grid, ham, kappa );  % initalization of the Gross-Pitaevskii object
\end{code}
Here \verb|grid| is the computational grid, \verb|ham| is the Hamiltonian consisting of the kinetic energy and the confinement potential, and \verb|kappa| is the nonlinearity.  Note that \verb!ham! must be a function or function handle that depends on the control parameter.  The main purpose of the \verb|grosspitaevskii| class is to allow for the simulation of the condensate time evolution within the framework of the Gross-Pitaevskii equation, using confinement potentials that can be controlled by some external control parameter (see Sec.~\ref{sec:control}).  Figure~\ref{fig:gpschematics} schematically depicts the solution scheme we are aiming at.

Let us consider the situation where a condensate initially resides in a one-dimensional harmonic-type well, and the confinement potential is subsequently transformed to a double well.  Such a setup has been analyzed in Ref.~\cite{hohenester.pra:07}.  In the demo program \verb|demogp1.m| we first set up the computational grid, the confinement potential, and the nonlinearity.

\begin{code}
units;                         %  length in micrometers, time in ms, hbar = 1
grid = grid1d( - 3, 3, 101 );  %  simulation grid
v = @( lambda ) ( lesanovsky1d( grid.x, lambda ) );  %  confinement potential
ham = @( lambda ) ( - 0.5 * grid.lap4 / ( 87 * mass ) + spdiag( v( lambda ) ) );
                               %  single particle Hamiltonian
kappa = pi;                    %  nonlinearity

tout = linspace( 0, 2, 100 );  %  output time
nt = length( tout );           %  number of time steps
lambda = control( tout, sqrt( min( 2 * tout / max( tout ), 1 ) ) );
                               %  control parameter
\end{code}
Note that the Hamiltonian must be a sparse matrix $H_{ij}$ in real space.  For that reason we put the confinement potential via \verb|spdiag| on the diagonal of the matrix.  In the last lines of the above demo program we have also specified the time interval $[0,2]$ of the simulation and have defined the time variation of the control parameter, that describes how the confinement potential is transformed from a single well to a double well.  With the chosen form the transformation occurs in the first half of the time interval, and the potential remains fixed in the second half of the interval. 

We next compute the Gross-Pitaevskii ground state using the optimal damping algorithm of Dion et al.~\cite{dion:07}.

\begin{code}
%  groundstate wavefunction
psi0 = groundstate( grosspitaevskii( grid, ham, kappa ), lambda( 0 ) );
%  plot ground state
plot( grid.x, psi0, 'b-', grid.x, 1e-3 * v( 0 ), 'r--' );
\end{code}
For double-well potentials the algorithm sometimes fails.  In these cases we recommend to use a more robust but less efficient algorithm based on state mixing by calling \verb|groundstate(...,'mix',1e-2)| instead.  Finally, we set up an ODE solver and solve the Gross-Pitaeskii equation for the control field $\lambda(t)$ in time

\begin{figure}
\centerline{\includegraphics[width=0.35\columnwidth]{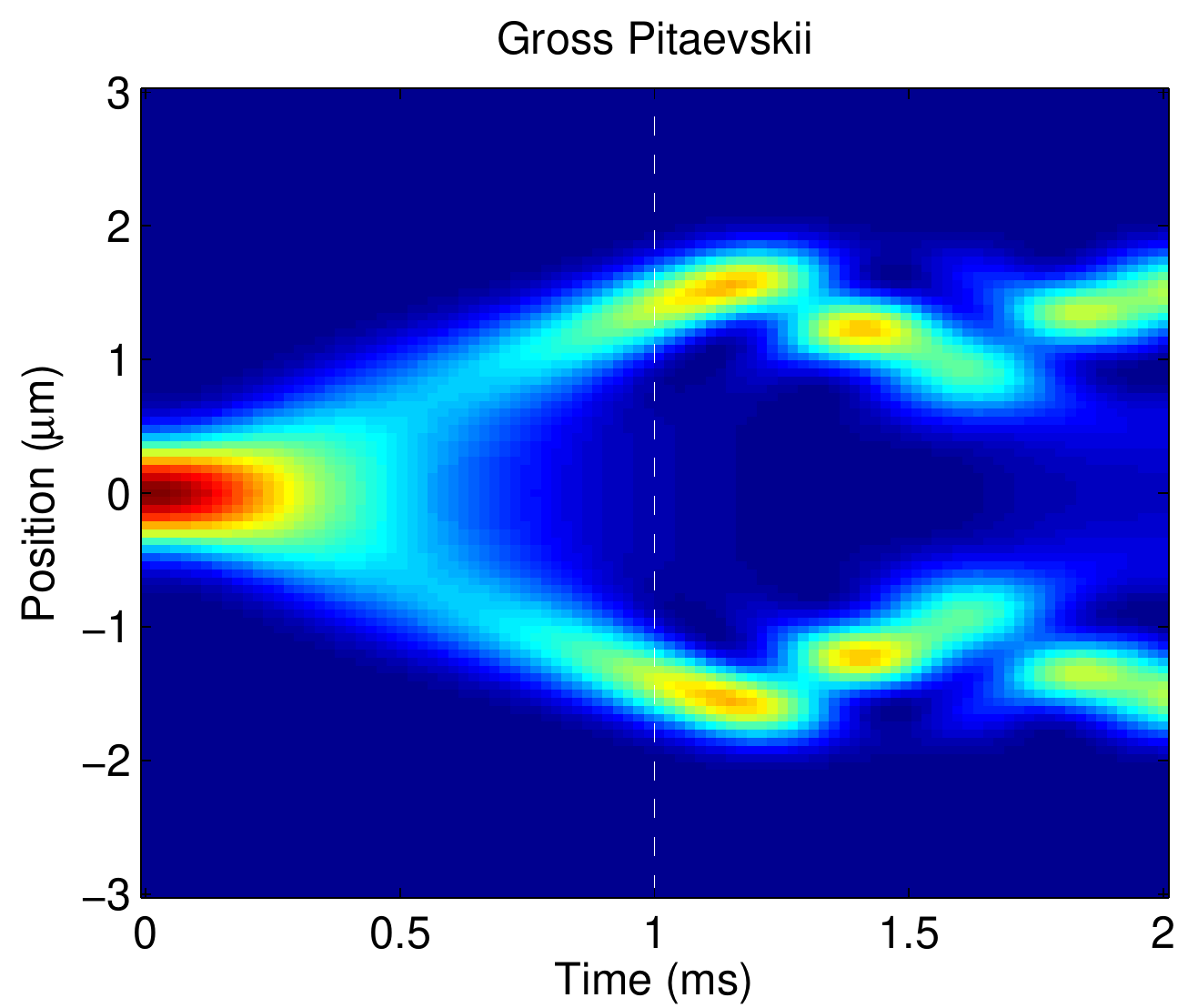}\quad
            \includegraphics[width=0.35\columnwidth]{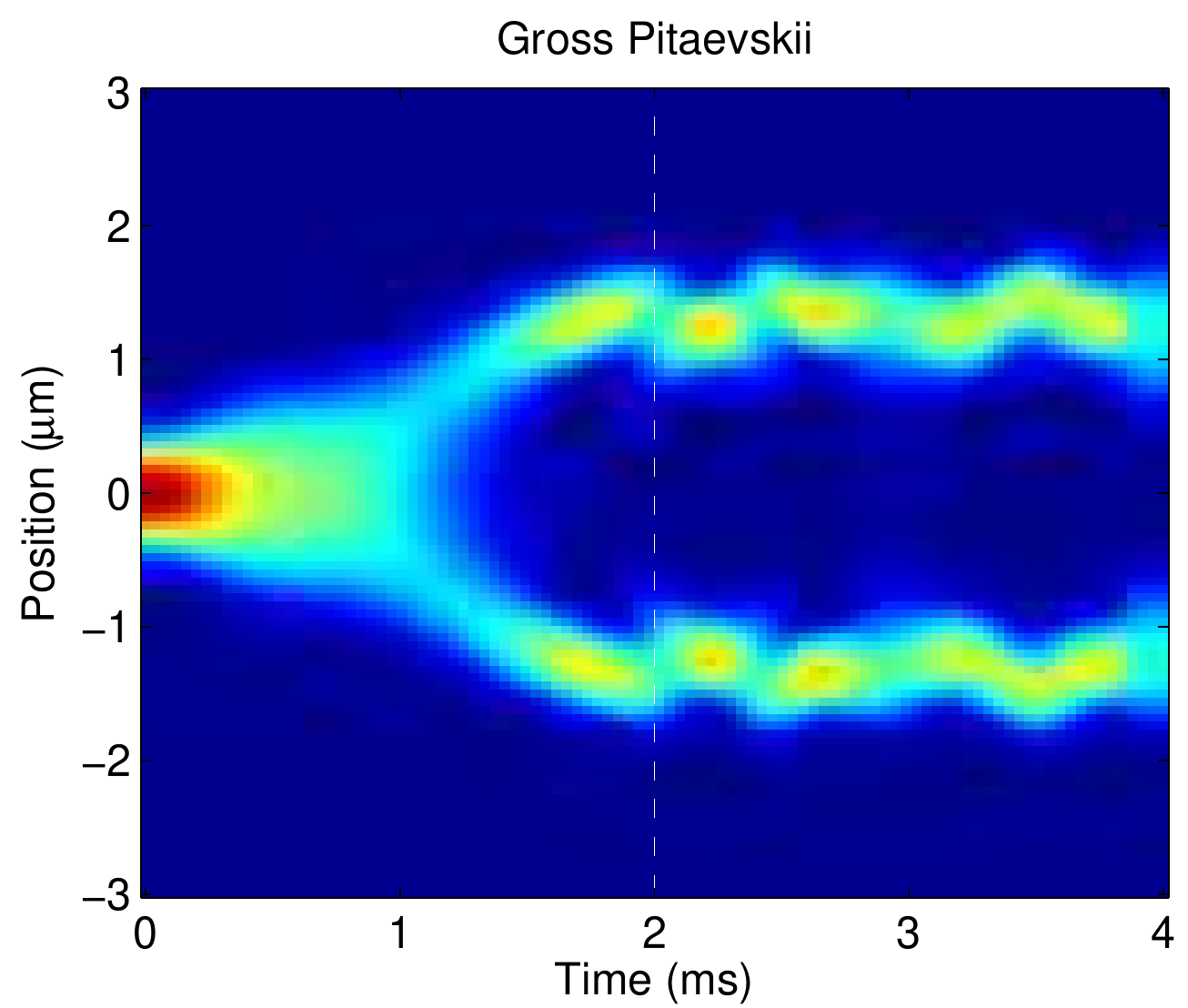}}
\caption{Density plot for time evolution of particle density for different splitting times.  The left (right) panel reports results for a splitting time of 1 ms (2 ms).  In the second half of the simulation the confinement potential is kept constant.}\label{fig:gpsplitden}
\end{figure}

\begin{code}
op = struct( 'nsub', 20, 'nout', 10, 'stepfun', 'ode23' );  %  options for ODE solver
[ psiout, tout ] = solve( psi0, tout, lambda, op );         %  solve differential equation

%  final output
imagesc( tout, grid.x, density( psiout ) );

xlabel( 'Time (ms)' );
ylabel( 'Position (\mum)' );
title( 'Gross Pitaevskii' );
\end{code}

The density plot in the left panel of Fig.~\ref{fig:gpsplitden} shows the time evolution of the particle density.  One sees that the condensate wavefunction splits and the density oscillates in the separated wells of the double-well potential.  If we increase the time of the splitting process by a factor of two, the splitting becomes more adiabatic and the wavefunction oscillates less, as shown in the right panel of the figure.  

The OCTBEC toolbox additionally allows for solutions of the Gross-Pitaevskii equation with the Crank-Nicolson or split operator techniques.  Details of this approach can be found in Ref.~\cite{hohenester.pra:07} and in \ref{sec:crank}.  For the Crank-Nicolson technique we simply have to change the options for the ODE solver

\begin{code}
%  options for ODE solver using the Crank-Nicolson scheme
op = struct( 'nsub', 2, 'nout', 10, 'stepfun', 'crank' );
\end{code}
The Crank-Nicolson scheme has the advantage that the norm of the wavefunction is always preserved and that one can typically use significantly larger time steps than with the Runge-Kutta technique. This is particularly advantageous for OCT calculations where the Gross-Pitaevskii equation has to be solved many times.  Note that we properly include the nonlinear term in our Crank-Nicolson approach, using a Newton iteration at each time step \cite{grond.pra:09b}, as described in more detail in \ref{sec:crank}.

To implement the split operator approach, we must equip the initial wavefunction with a kinetic energy operator evaluated in wavenumber space, and a confinement potential operator evaluated in real space.

\begin{code}
%  initialize split operator
psi0.t = @( lambda ) ( - 0.5 * grid.ilap / ( 87 * mass ) );
psi0.v = @( lambda ) ( v( lambda ) );
%  options for ODE solver using the Crank-Nicolson scheme
op = struct( 'nsub', 2, 'nout', 10, 'stepfun', 'split' );
\end{code}
Again the norm of the wavefunction is always preserved and one can use significantly larger time steps in comparison to the Runge-Kutta scheme.  To access the wavefunction and the particle density of a \verb|grosspitaevskii| object \verb|psi|, we can use the following commands.

\begin{code}
orb = orbital( psi );  %  or double( psi ), Gross-Pitaevskii wavefunction
den = density( psi );  %  particle density for Gross-Pitaevskii wavefunction
\end{code}
Note that the above commands also work for cell arrays as returned from the ODE solvers.

\subsection{Multi-configurational Hartree method for bosons (MCTDHB)}\label{sec:mctdhb}

The multi-configurational Hartree method for bosons (MCTDHB) has been developed in recent years by Cederbaum, Alon, and coworkers \cite{alon:08}.  The main idea is to provide several \textit{orbitals} $\phi_i(\bm r)$, which are determined in a self-consistent fashion, and to distribute the atoms among these orbitals.  The total wavefunction for $m$ orbitals is then of the form
\begin{equation}\label{eq:mctdhbwavefunction}
\Psi=\sum_{i_1}\sum_{i_2}\dots\sum_{i_m} C_{i_1i_2\dots i_m}
    \Bigl(a_1^\dagger\Bigr)^{i_1}
    \Bigl(a_2^\dagger\Bigr)^{i_2} \dots
    \Bigl(a_m^\dagger\Bigr)^{i_m} |{\rm vac}\rangle\,,
\end{equation}
where $i_1+i_2+\dots=N$ gives the total number of bosons and $C_{i_1,i_2,\dots}$ characterizes the distribution of atoms between the orbitals.  The dynamics of the orbitals $\phi_i(r,t)$ and the atom-number part $C$ of the wavefunction are obtained from a variational principle that choses the orbitals in an ``optimal way'', as discussed in some length in Ref.~\cite{alon:08}.  The working equation for the orbitals then becomes
\begin{equation}\label{eq:mctdhborb}
  i\dot\phi_i=\mathcal{P}\left[\left(-\frac{\hbar^2\nabla^2}{2M} + V(\bm r,\lambda(t))
  \right)\phi_i + 
  \kappa\sum_{jklm}\rho_{im}^{-1}\rho_{mjkl}^{(2)}\phi_j^*\phi_k\phi_l\right]\,,
\end{equation}
where $\rho$ and $\rho^{(2)}$ denote the one- and two-particle density matrix, respectively, and  the projector $\mathcal{P}=1-\sum_i|\phi_i\rangle\langle\phi_i|$ assures that the ensuing term is orthogonal to the orbitals~\cite{alon:08}.  The atom number part is governed by the Schr\"odinger equation $i\dot C=HC$, with the many-body Hamiltonian $H$, which has to be solved in parallel with the orbital part of Eq.~\eqref{eq:mctdhborb}.

Within the OCTBEC toolbox, we provide an implementation with $m$ orbitals, although we have primarily used and tested two orbitals.  The MCTDHB wavefunction is initialized with 
\begin{code}
psi = mctdhb( grid, ham, kappa, m, n );  % initialization of MCTDHB wavefunction
\end{code}
Here \verb|grid| is the computational grid, \verb|ham| is the single-particle Hamiltonian, \verb|kappa| is the nonlinear atom-atom interaction, \verb|m| is the number of orbitals, and \verb|n| is the number of atoms.  Upon initialization, a \verb|mctdhb| object with the following properties is created:

\medskip
\begin{tabular}{ll}
\texttt{orb} & orbital part of wavefunction, \\
\texttt{num} & atom number part of wavefunction.\\
\end{tabular}
\medskip

\begin{figure}
\centerline{
  \includegraphics[width=0.3\columnwidth]{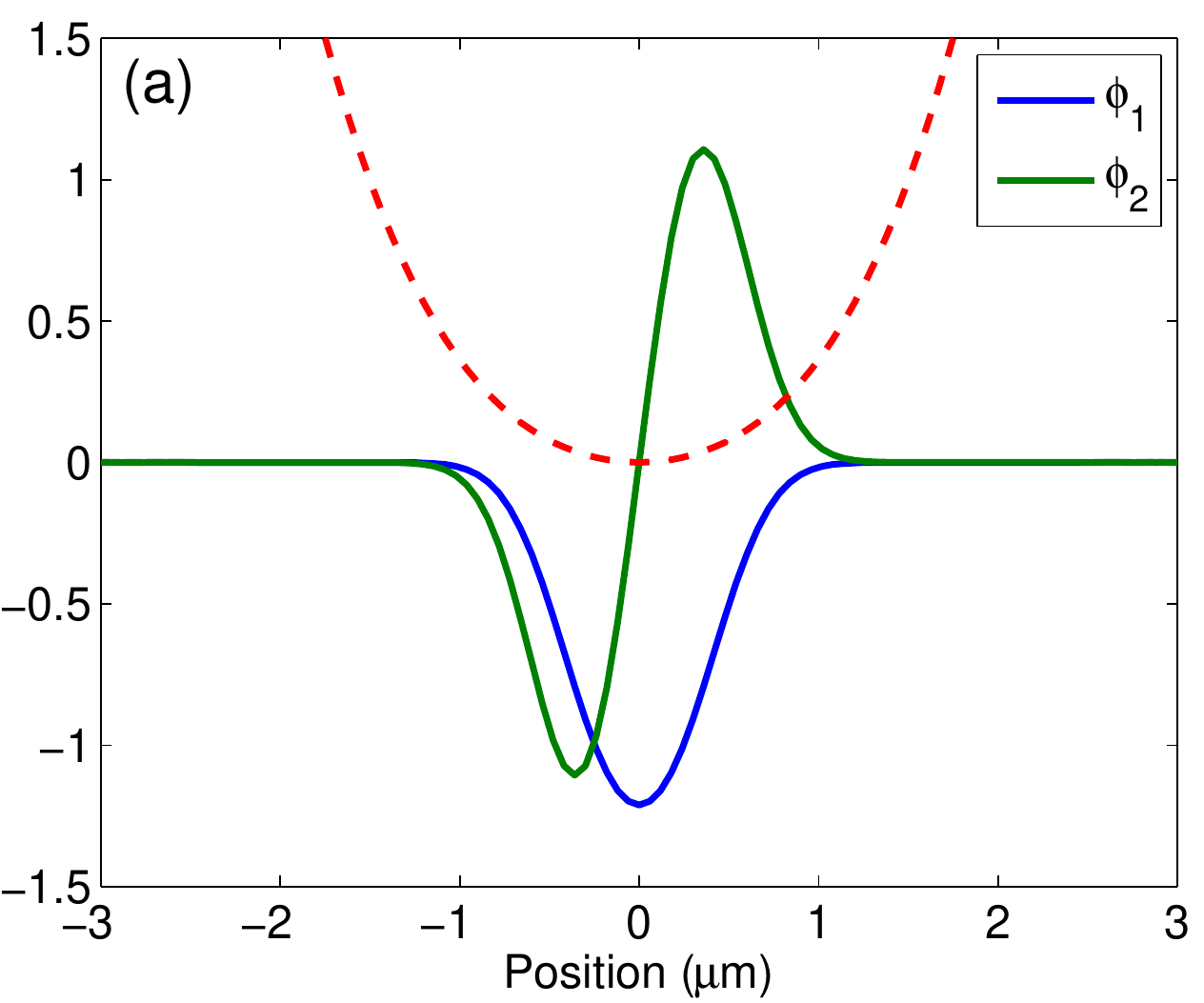}\quad
  \includegraphics[width=0.3\columnwidth]{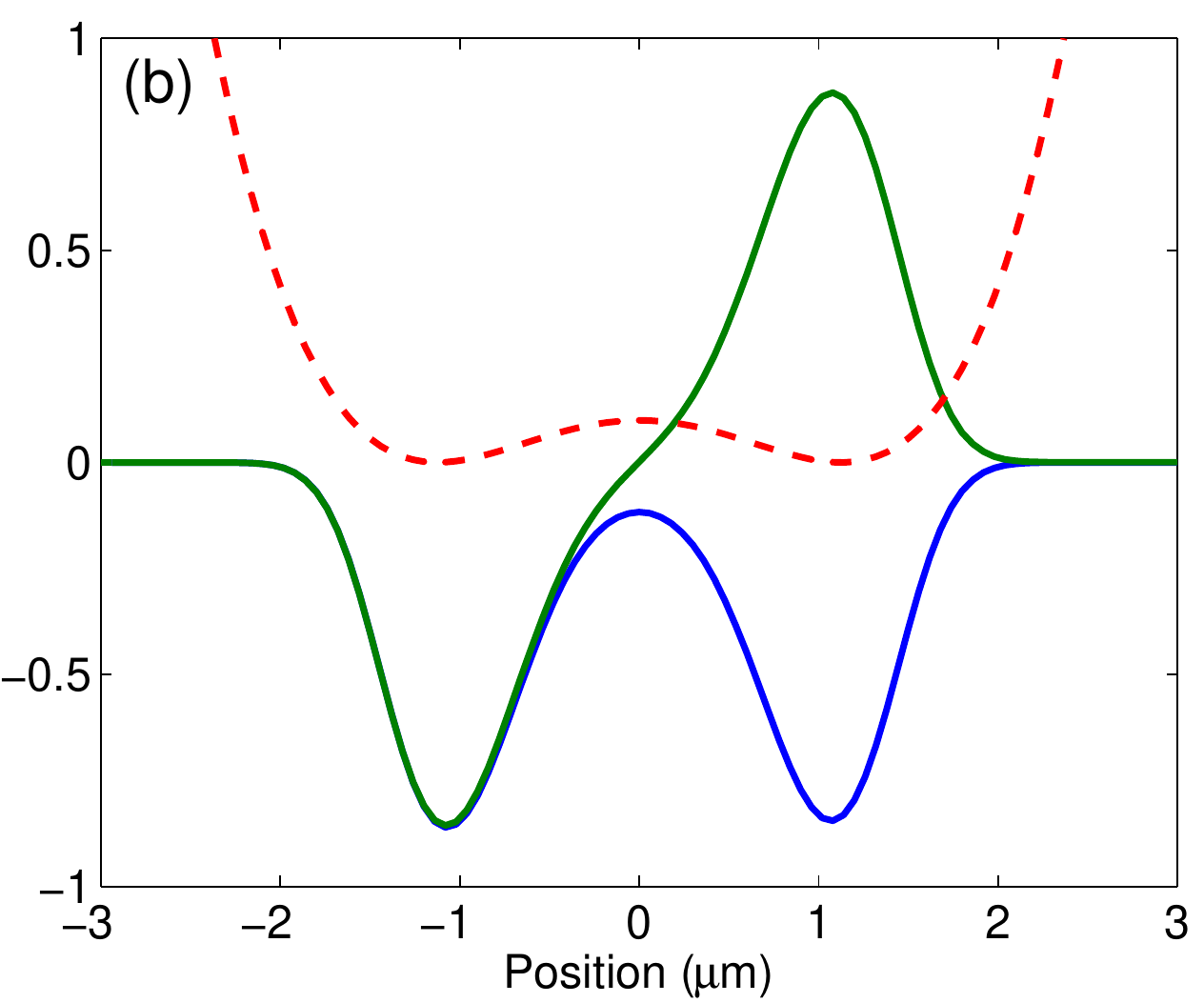}\quad
  \includegraphics[width=0.3\columnwidth]{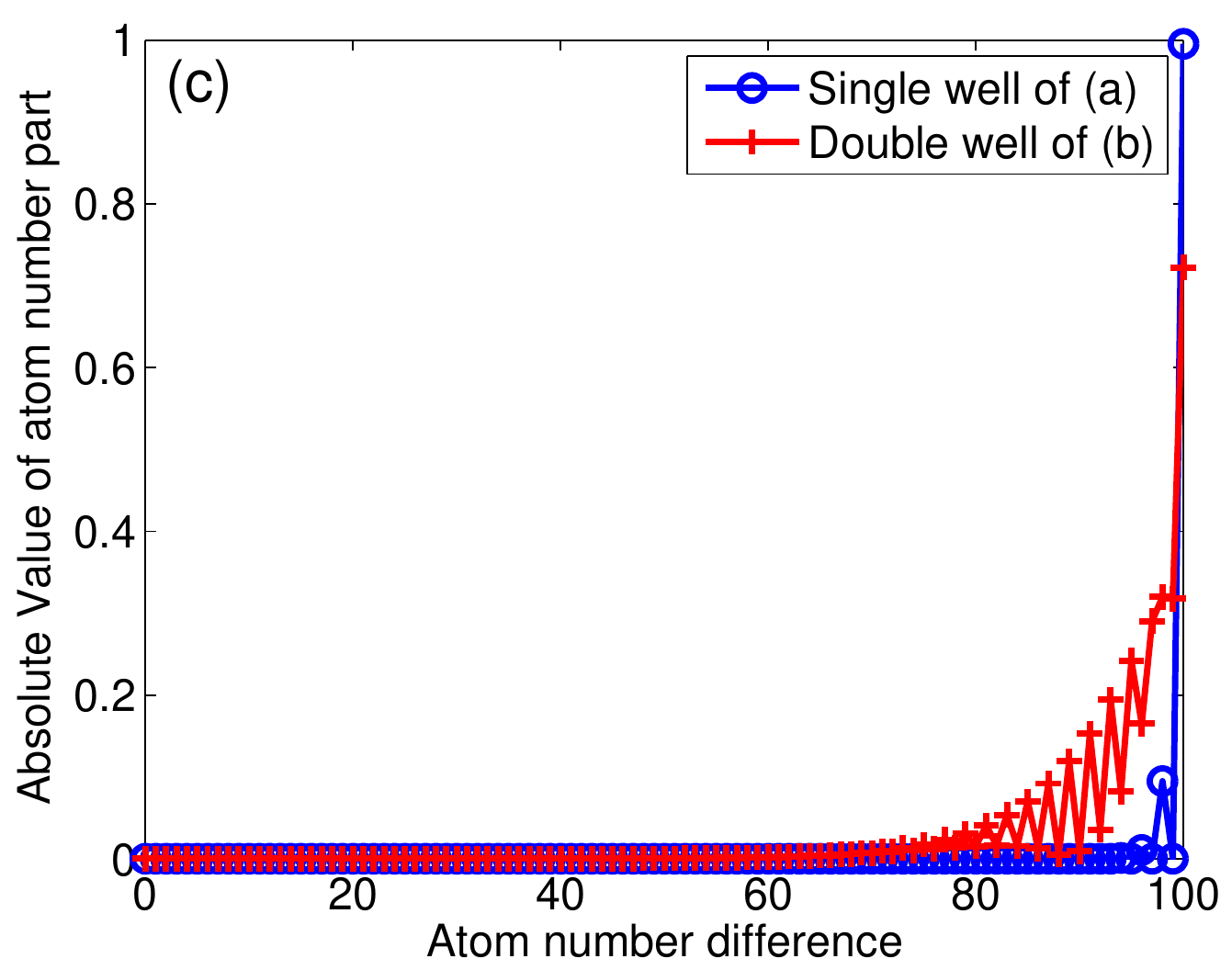}}
\caption{MCTDHB groundstate orbitals for (a) single and (b) double well potential.  The dashed lines report the confinement potential.  (c) Atom number part $C_{i,N-i}$ of wavefunction.  For the single well only the orbital with gerade symmetry is significantly populated, wheres for the split trap both orbitals of gerade and ungerade symmetry become populated, thus indicating condensate fragmentation.}\label{fig:mctdhb1}
\end{figure}

\noindent Let us consider a simple example of condensate splitting by transforming the confinement potential from a single to a double well, as shown in the demo program \texttt{demomctdhb1.m}.  We first set up a computational grid and define a single-particle Hamiltonian with a confinement potential that can be modified through a control parameter.  These steps are very similar to the Gross-Pitaevskii simulations previosuly discussed.
\begin{code}
units;                         %  length in micrometers, time in ms, hbar = 1
grid = grid1d( - 3, 3, 101 );  %  set up computational grid
v = @( lambda ) ( lesanovsky1d( grid.x, lambda ) );  %  confinement potential
ham = @( lambda ) ( - 0.5 * grid.lap4 / ( 87 * mass ) + spdiag( v( lambda ) ) );
                               %  single particle Hamiltonian

n = 100;                           %  number of particles
kappa = 0.5 * 2 * pi / ( n - 1 );  %  nonlinearity (500 Hz)
m = 2;                             %  number of orbitals
psi0 = groundstate( mctdhb( grid, ham, kappa, m, n ), 0 );
                                   %  MCTDHB groundstate

%  plot orbitals and C vector
subplot( 1, 2, 1 );  plot( grid.x, psi0.orb, grid.x, 1e-2 * v( 0 ), 'r--' );
subplot( 1, 2, 2 );  plot( abs( psi0.num ), 'o' );
\end{code}
In the last lines we plot the orbitals for the MCTDHB groundstate, which exhibit gerade and ungerade symmetry, as well as the distribution of atoms among these orbitals (see Fig.~\ref{fig:mctdhb1}).  For the single well practically all atoms reside in the orbital with gerade symmetry, and the occupation of the ungerade orbital is very small (corresponding to the situation of an undepleted condensate).  The situation changes dramatically for a double well potential, as shown in Fig.~\ref{fig:mctdhb1}.
\begin{code}
%  MCTDHB groundstate for initial double well potential
psi0 = groundstate( mctdhb( grid, ham, kappa, m, n ), 0.85 );
\end{code}
Here a much larger fraction of the atoms populates the first excited orbital of ungerade symmetry, corresponding to condensate fragmentation.  As previously discussed for the Gross-Pitaevskii groundstate, for further splitting one should use \verb|psi0=groundstate(...'mix',1e-2)|.

\begin{figure}
\centerline{
  \includegraphics[width=0.3\columnwidth]{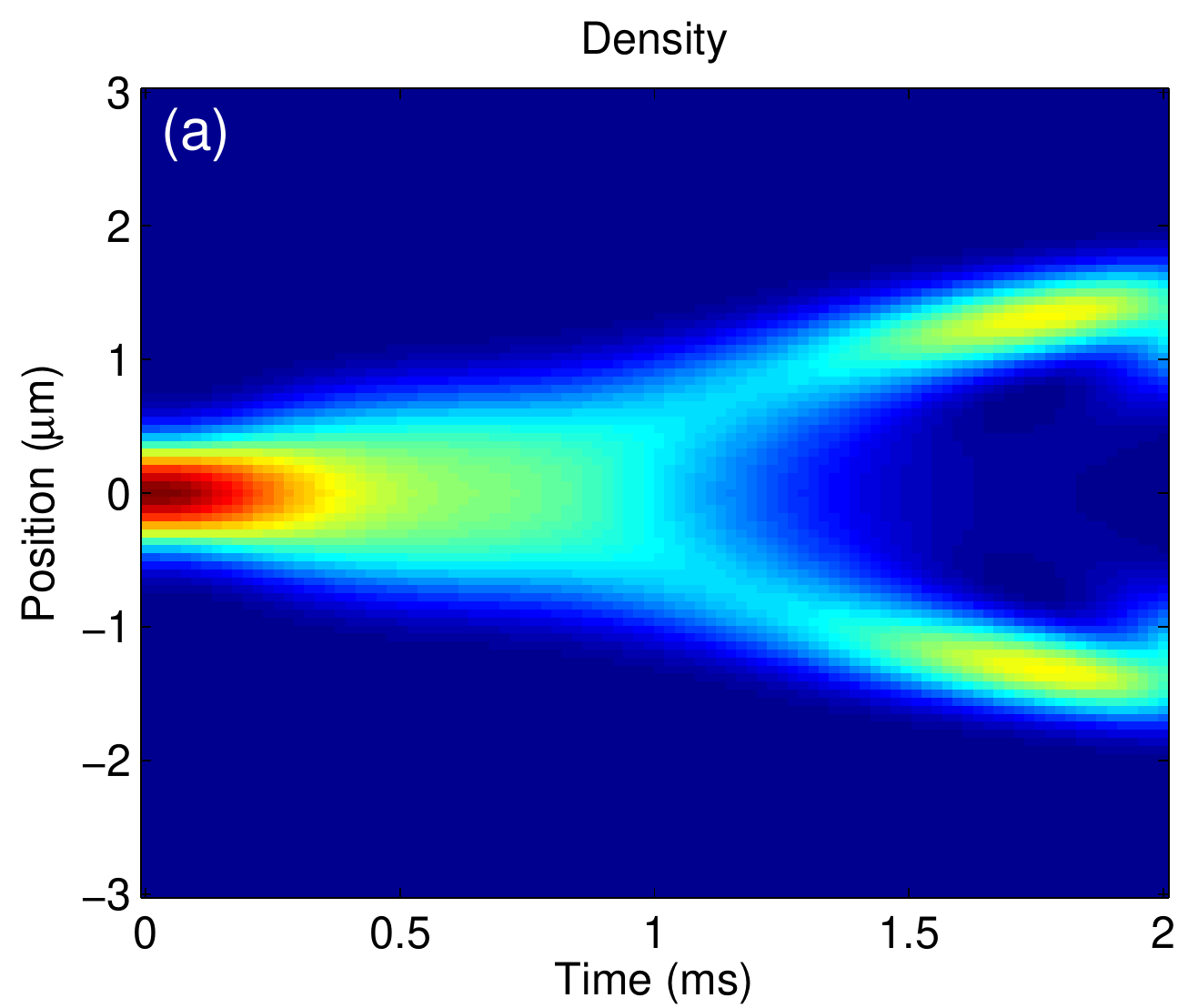}\quad
  \includegraphics[width=0.3\columnwidth]{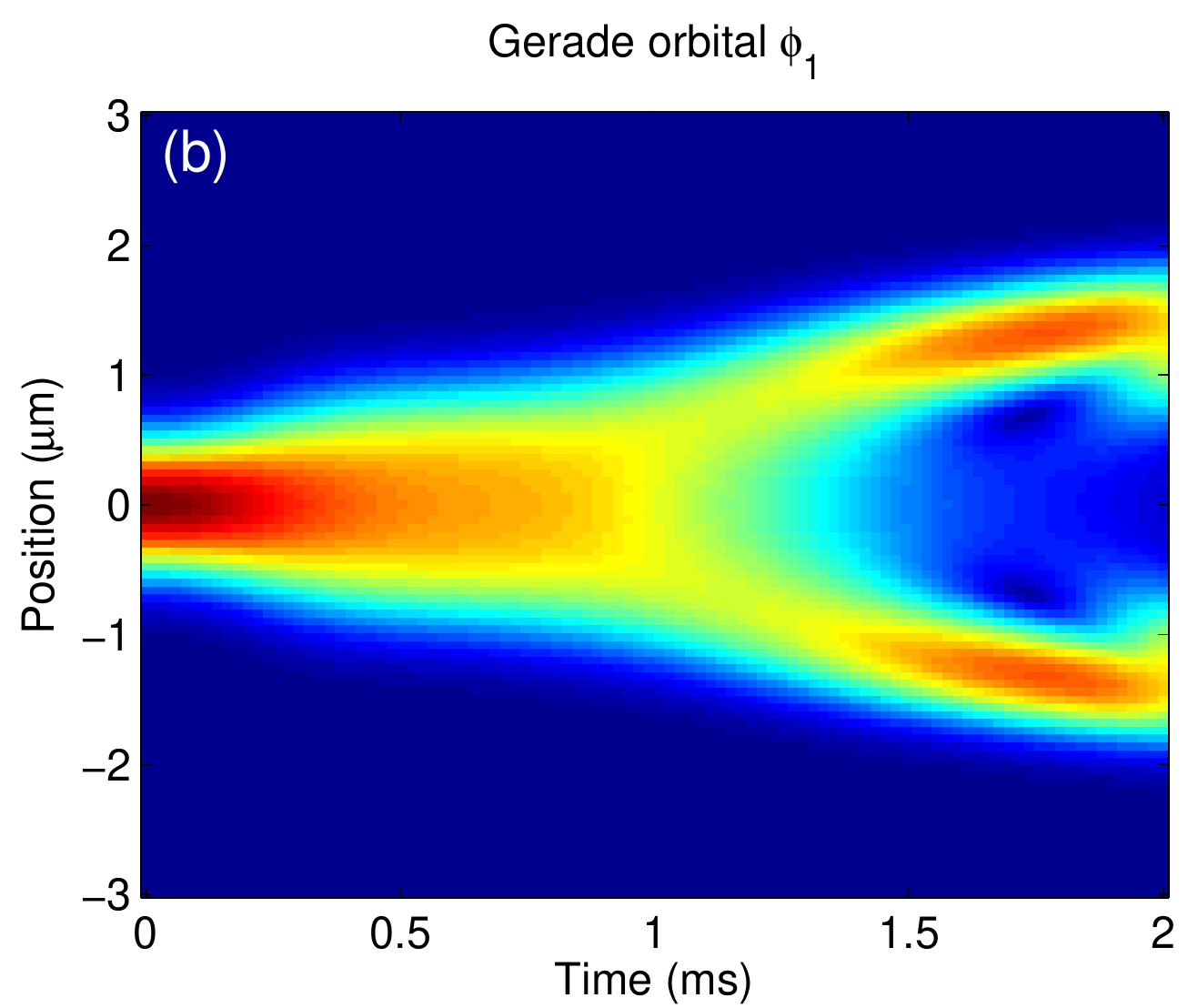}\quad
  \includegraphics[width=0.3\columnwidth]{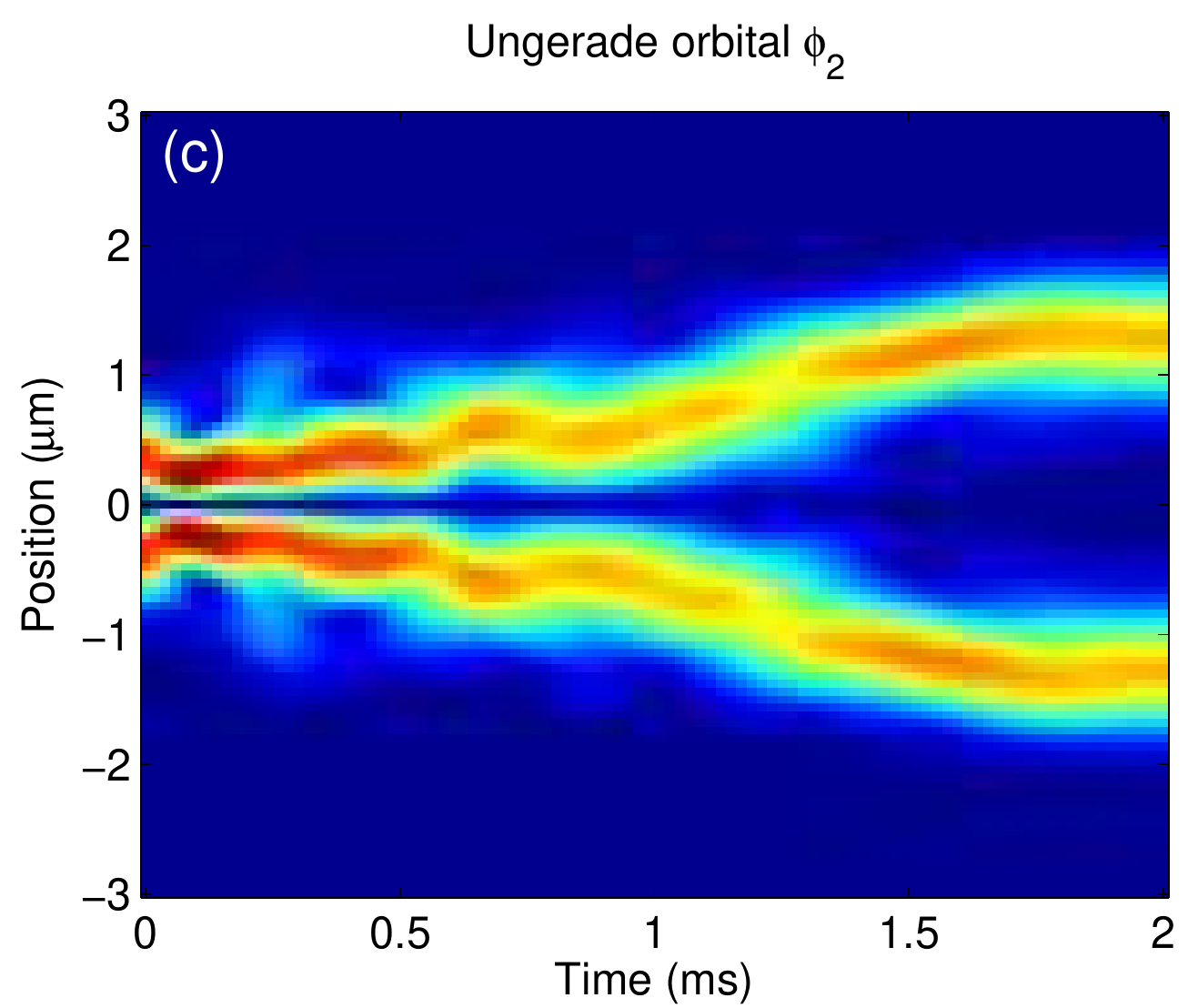}}
\caption{Splitting process of MCTDHB wavefunction.  Time evolution of (a) particle density, as well as of orbitals with (b) gerade and (c) ungerade symmetry.  The atom number part $C$ of the wavefunction (not shown) determines how the atoms become distributed between these orbitals.
}\label{fig:mctdhb2}
\end{figure}

We next compute the time evolution of a system that initially starts in the groundstate of a single-well potential, and the potential is subsequently transformed to a double well.
\begin{code}
tout = linspace( 0, 3, 100 );  %  output times
nt = length( tout );           %  number of time steps
lambda = control( tout, sqrt( tout / max( tout ) ) );      %  control parameter 

op = struct( 'nsub', 20, 'nout', 5, 'stepfun', 'ode23' );  %  options for ODE solver
[ psiout, tout ] = solve( psi0, tout, lambda, op );        %  solve MCTDHB equations

%  final output
imagesc( tout, grid.x, density( psiout ) );

xlabel( 'Time (ms)' );
ylabel( 'Position (\mum)' );
title( 'MCTDHB(2)' );
\end{code}
We can also plot the time evolution of the two orbitals, as shown in Fig.~\ref{fig:mctdhb2}.

\begin{code}
%  plot time evolution of orbitals
subplot( 1, 2, 1 );  imagesc( tout, grid.x, abs( orbital( psiout, 1 ) ) );
subplot( 1, 2, 2 );  imagesc( tout, grid.x, abs( orbital( psiout, 2 ) ) );
\end{code}

\noindent The number part of the MCTDHB wavefunction is often hard to interpret because atoms are distributed among time-varying orbitals.  Plotting of the time dependence of $C$ can be done through
\begin{code}
plot( tout, abs( num( psiout{ 1 } ) ) );  %  plot modulus of C vector
\end{code}

The \verb|OCTBEC| toolbox additionally allows for solutions of the MCTDHB equation with the Crank-Nicolson technique.  Details of this approach can be found in Ref.~\cite{grond.pra:09b} as well as in \ref{sec:crank}.  For the Crank-Nicolson technique we simply have to change the options for the ODE solver
\begin{code}
%  options for ODE solver using the Crank-Nicolson scheme
op = struct( 'nsub', 2, 'nout', 10, 'stepfun', 'crank' );
\end{code}
The Crank-Nicolson scheme has the advantage that the norm of the wavefunction is always preserved and that one can typically use significantly larger time steps than with the Runge-Kutta technique. This is particularly advantageous for OCT calculations where the MCTDHB equations have to be solved many times.  Finally, for the calculation of the true MCTDHB groundstate we refer to the help pages and to the demo program \verb!demogroundstatemctdhb.m !.

\subsection{Generic few-mode models}\label{sec:fewmode}

The OCTBEC toolbox provides several classes for the description of the atom number part, where atoms become distributed between static or time-dependent orbitals, and the time evolution is governed by a Hamiltonian matrix.  The most general classes are \verb|fewmodepair| and \verb|fockstate|, which we will describe first.  The classes \verb|twomodepair| and \verb|twomodespin| are more specialized and can be employed for the solution of a two-mode model, that has a longstanding history in the description of BECs \cite{milburn:97,javanainen:99}.

Consider a system of \verb|n| atoms that reside in \verb|m| orbitals.  The wavefunction can be expanded in the atom number Hilbert space, the Fock space, according to Eq.~\eqref{eq:mctdhbwavefunction}.  To set up the Fock space, we call
\begin{code}
spin = fewmodepair( n, m, cutoff );  %  set up Fock space for n atoms and m orbitals
\end{code}
Here \verb|cutoff| is an optional vector that determines the maximal number of atoms within a given orbital.  For instance, we obtain
\begin{code}
spin = fewmodepair( 100, 2 )   %  set up Fock space for 100 atoms in 2 orbitals

fewmodepair :
        n: 100
        j: {2x2 cell}
    state: [101x2 double]
\end{code}
where \verb|j{k,l}| denotes the pseudospin operator $J_{kl}=\hat a_k^\dagger\hat a_l$ and \verb|state| is a matrix for all possible atom number configurations.  The size of the Fock space is given by \verb|size(spin.state,1)|.

\begin{figure}
\begin{pdffigure}
\centerline{\includegraphics[width=0.35\columnwidth]{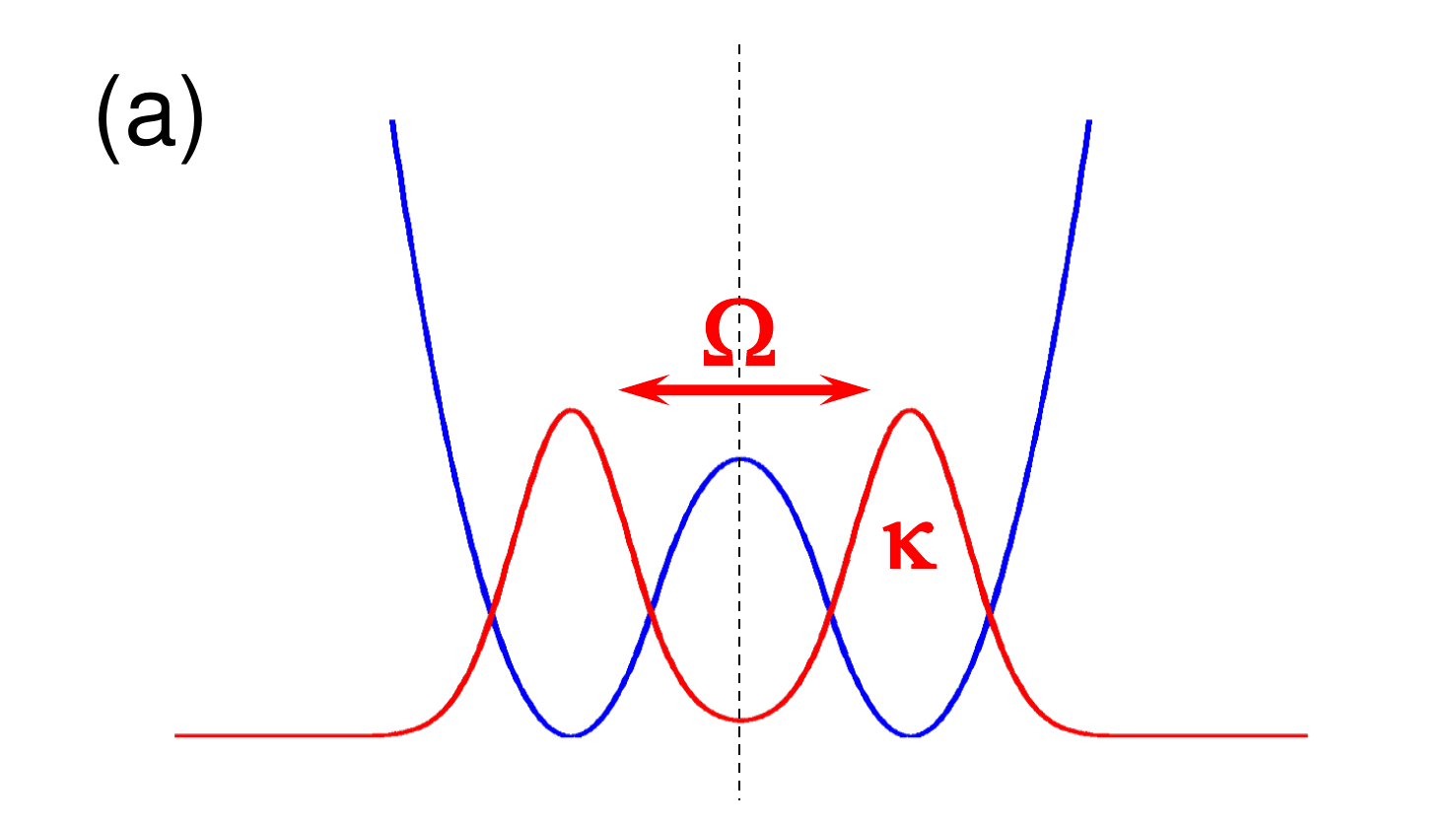}\quad
            \includegraphics[width=0.30\columnwidth]{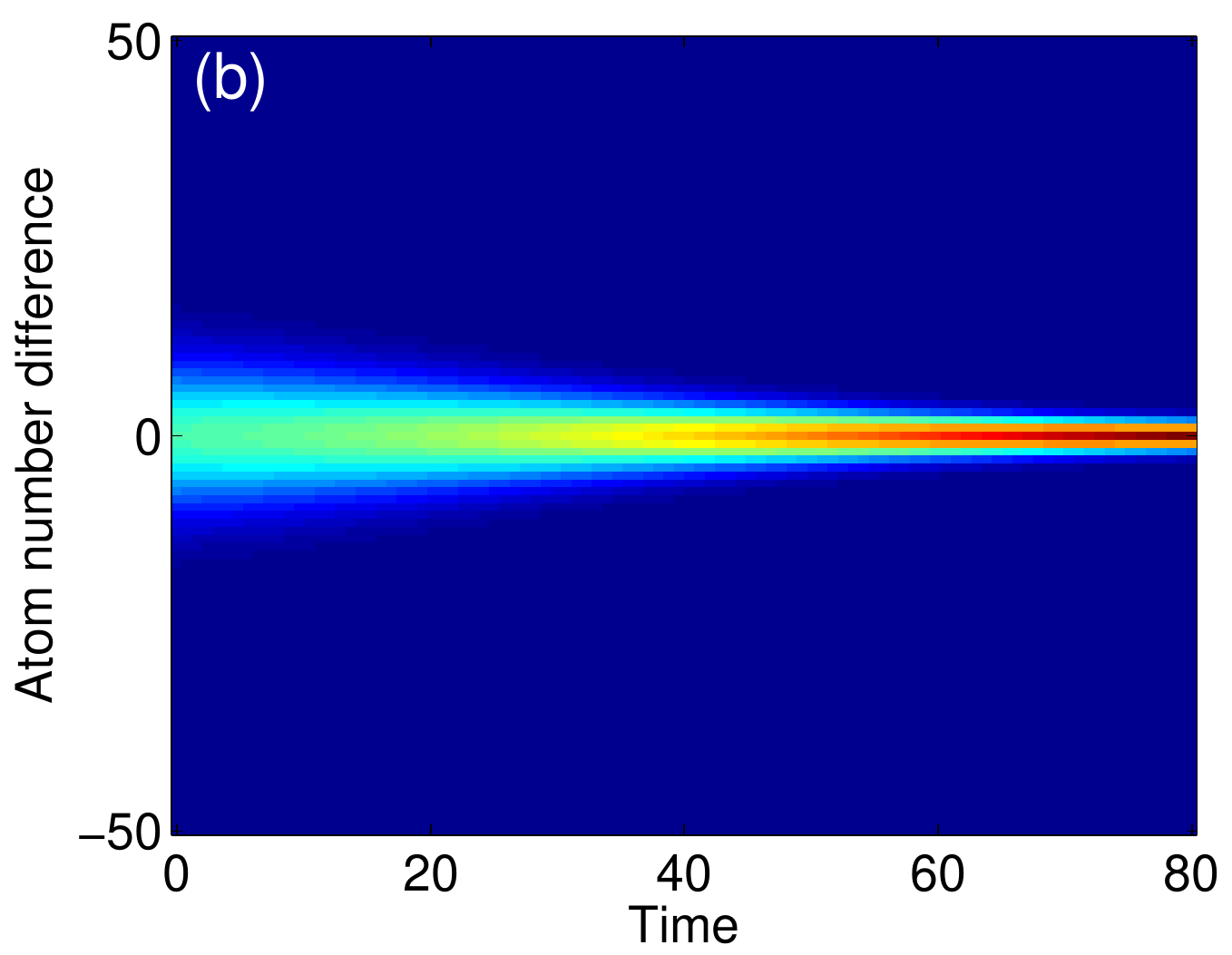}}
\end{pdffigure}
\caption{(a) Schematics of two-mode model.  The atoms reside in two orbitals (left and right), and are coupled by a tunneling element $\Omega$.  By turning off the tunnel coupling, the atom number distribution becomes squeezed because of the nonlinear atom-atom interaction $\kappa$.  (b) Time evolution of atom number distribution for an exponential decrease of $\Omega$.}\label{fig:fockstate}
\end{figure}

We next consider the two-mode model discussed by Javanainen and Ivanov~\cite{javanainen:99}, where atoms reside in the left or right well of a double-well potential, subject to tunneling and nonlinear interactions, as shown in Fig.~\ref{fig:fockstate}(a).  In the figure the blue line represents the double-well potential and the red line the orbital part of the wavefunction.  The Hamilton describing the dynamics of this two-mode model is of the form
\begin{equation}\label{eq:fewmode}
 H=-\frac\Omega 2\left(\hat a_1^\dagger \hat a_2+\hat a_2^\dagger \hat a_1\right)+
    \kappa\left(\hat a_1^\dagger \hat a_1^\dagger \hat a_1\hat a_1+
                \hat a_2^\dagger \hat a_2^\dagger \hat a_2\hat a_2\right)\,.
\end{equation}
The first term describes tunneling between two wells, with the tunnel coupling $\Omega$, and $\kappa$ is the nonlinear atom-atom interaction.  The demo program \verb!demofewmodepair.m! shows the
implementation of such a two-mode model.
\begin{code}
n = 100;        %  number of particles
kappa = 1 / n;  %  nonlinearity parameter
spin = fewmodepair( n, 2 );  %  peudospin object

%  tunneling and nonlinear coupling
tun = - 0.5 * [ 0, 1; 1, 0 ];
non = kappa * accumarray( [ 1, 1, 1, 1; 2, 2, 2, 2 ], 1 );
%  transform to operators
optun = spin.op( tun );
opnon = spin.op( non );

ham = @( lambda ) ( lambda * optun + opnon  );  %  Hamiltonian function
\end{code}
The operators, defined through the \verb|fewmodepair| class, work together with \verb|fockstate| objects that define a wavefunction within the Fock space.  A \verb|fockstate| object is initialized with 
\begin{code}
psi = fockstate( spin, ham );  %  initialize FOCKSTATE object
\end{code}
Here \verb|spin| is for instance the \verb|fewmodepair| object defined above, which defines the Hilbert space in the atom number basis, and \verb|ham| is a Hamiltonian acting in this Fock space.  We assume that the Hamiltonian can be modified by some external control parameter.  Upon initialization, the \verb|fockstate| object has the following properties

\medskip
\begin{tabular}{ll}
\texttt{ham} & Hamiltonian in atom-number basis, \\
\texttt{num} & wavefunction in atom number space,\\
\texttt{spin} & pseudospin object.\\
\end{tabular}
\medskip

\noindent The combination of the \verb|fewmodepair| and \verb|fockstate| classes allows to
solve time-dependent problems in a quite elegant fashion.  Consider the situation where the system is initially in the ground state governed by tunneling, and the tunneling is turned off exponentially at later times.
\begin{code}
tout = linspace( 0, 80, 100 );                     %  output times
lambda = control( tout, 3 * exp( - tout / 10 ) );  %  tunnel coupling
[ psi0, ene ] = groundstate( fockstate( spin, ham ), lambda( 0 ) );  %  initial wavefunction

op = struct( 'nsub', 100, 'nout', 10, 'stepfun', 'runge4' );  %  Runge-Kutta integration
[ psiout, tout ] = solve( psi0, tout, lambda, op );           %  solve differential equation

%  density map of atom number part
imagesc( tout, 0.5 * n * [ - 1, 1 ], abs( num( psiout ) ) );

xlabel( 'Time' );
ylabel( 'Atom number difference' );
\end{code}
The simulation results is shown in Fig.~\ref{fig:fockstate}(b).  One sees that the atom number fluctuations around the mean value of $n/2$ are initially relatively large, corresponding to a binomial state, and become significantly reduced at later times owing to the nonlinear atom-atom interaction.

\begin{figure}
\begin{pdffigure}
\centerline{\includegraphics[width=0.8\columnwidth]{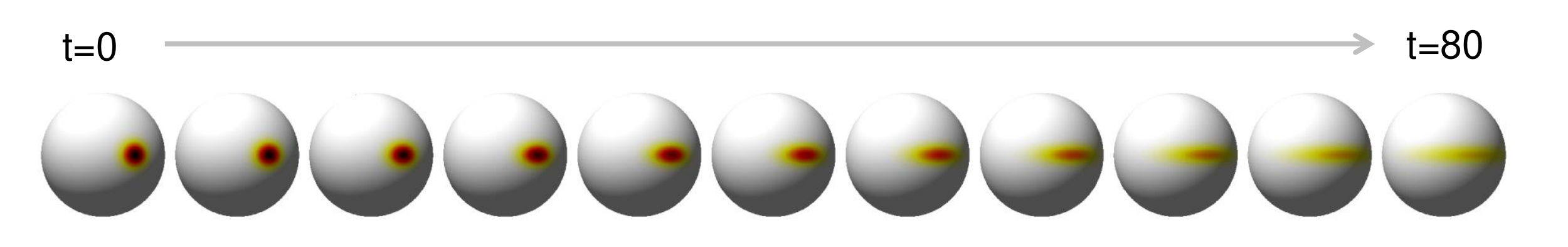}}
\end{pdffigure}
\caption{Time evolution of number squeezing, through exponential turning-off of the tunnel coupling, as visualized on Bloch sphere.  See \texttt{demofewmodepairbloch.m} for the corresponding simulation.}\label{fig:blochsphere}
\end{figure}

To use a Crank-Nicolson solution scheme, we have to replace the options for the ODE solver with
\begin{code}
%  options for Crank-Nicolson integration
op = struct( 'nsub', 10, 'nout', 10, 'stepfun', 'crank' );
\end{code}
When psiout is a cell array of \verb|fockstate| objects, as usually returned from the ODE solver, the map of atom number wavefunctions can be obtained through
\begin{code}
psi = num( psiout );  %  wavefunction map from cell array PSIOUT
\end{code}
For a two-mode model the wavefunction in the atom number space can be conveniently plotted on the Bloch sphere, as discussed by Arecchi et al. \cite{arecchi:72}.  Within the OCTBEC toolbox we set up the Bloch sphere with
\begin{code}
% default initialization of Bloch sphere 
sph = blochsphere( n );
%  initialization of Bloch sphere with user-defined discretization
sph = blochsphere( n, nsph );
\end{code}
Here \verb|n| is the number of atoms and \verb|nsph| is an optional parameter (default value 200) that controls the discretization of the Bloch sphere.  Through \verb!plot(sph,num(psiout{1}))! we next plot the wavefunction on the Bloch sphere.  Fig.~\ref{fig:blochsphere} show the time evolution of the splitting process on the Bloch sphere, as computed with the demo program \verb!demofewmodepairbloch.m!. Initially the system is in a binomial state (left) and number squeezing occurs at the end of the splitting process (right).  
The OCTBEC toolbox additionally provides a class \verb|twomodepair|, that is very similar to the \verb|fewmodepair| class for two orbitals, as well as a \verb|twomodespin| class that uses pseudospin operators introduced e.g. by Milburn et al.~\cite{milburn:97}.  Details can be found in the help pages and the demo files.

\section{Optimal quantum control}\label{sec:oct}

\subsection{Optimal control theory}

The main purpose of the OCTBEC toolbox is to submit the different model systems to optimal control theory (OCT).  In this section we provide the theoretical background for the OCT framework.  A more detailed discussion can be found in Refs.~\cite{hohenester.pra:07,hohenester.fdp:09,buecker:13}.  In the following we first discuss optimization for a Gross-Pitaevskii wavefunction.  The general OCT framework will be discussed at the end.  For the working equations of the MCTDHB model, we refer the interested reader to Ref.~\cite{grond.pra:09b} and \ref{sec:eqsoptmctdhb}.

Consider the example \verb!demogp1.m! for condensate splitting within the Gross-Pitaevskii framework, discussed in Sec.~\ref{sec:gp}, where after splitting the wavefunction oscillates around the minima of the split trap (Fig.~\ref{fig:gpsplitden}).  In what follows we are seeking for a time variation of the control parameter $\lambda(t)$ such that the condensate wavefunction remains at rest after the splitting process.  This task can be accomplished with optimal control theory.  Let $\psi_d(x)$ denote the \textit{desired} groundstate wavefunction of the splitting process, which we can compute through
\begin{code}
%  desired wavefunction at terminal time
psid = groundstate( grosspitaevskii( grid, ham, kappa ), lambda.last, 'mix', 1e-2 );
\end{code}
OCT determines the optimal control $\lambda_{\rm opt}(t)$ such that the system is brought from the initial state $\psi_0(x)$ to the desired state $\psi_d(x)$ in an optimized fashion.  First, we introduce a \textit{cost function} $J(\psi,\lambda)$ that measures the success of a given control $\lambda(t)$, e.g., through
\begin{equation}\label{eq:costtrap}
  J(\psi,\lambda)=\frac 12\|\psi(T)-\psi_d\|^2+
  \frac\gamma 2\int_0^T \left(\dot\lambda(t)\right)^2\,dt\,. 
\end{equation}
The first term on the right-hand side becomes minimal when the terminal wavefunction $\psi(T)$ matches the desired state.  The second term penalizes strong variations of the control parameter and is needed to make the OCT problem well posed.  Through $\gamma$ it is possible to weight the importance of wavefunction matching and control smoothness, and in most cases we set $\gamma\ll 1$.  The optimal control problem under consideration can now be written as
\begin{equation}
  \mbox{min}\, J(\psi,\lambda)\quad\mbox{subject to}\quad
   i\dot\psi=\bigl(H_\lambda+\kappa|\psi|^2\bigr)\psi\,,\,\,
   \psi(0)=\psi_0\,.
\end{equation}
Note that we have explicity indicated the $\lambda$-dependence of the single-particle Hamiltonian.  The above equation states that we are looking for an optimal control that minimizes the cost function.  But in order to bring the system from the initial state $\psi_0$ to the terminal
state $\psi(T)$ we have to fulfill the Gross-Pitaevskii equation, which enters as a \textit{constraint} in our optimization problem.

The constrained optimization problem can be turned into an unconstrained one by means of Lagrange multipliers $p(t)$.  To this end, we introduce a Lagrange function
\begin{equation}
  L(\psi,p,\lambda)=J(\psi,\lambda)+\mbox{Re}\Biggl[
  \int_0^T \Bigl< p(t)\Bigr|i\dot\psi(t)-
  \left(H_\lambda+\kappa|\psi(t)|^2\right)\psi(t)\Bigr> dt\Biggr]\,. 
\end{equation}
The Lagrange function has a saddle point at the minimum of $J(\psi,\lambda)$ where all derivatives $\delta L/\delta\psi^*$, $\delta L/\delta p^*$, and $\delta L/\delta\lambda$ become zero.  Performing
functional derivatives in the Lagrange function, we then arrive at the following set of equations
\begin{align}
   i\dot\psi&=\left(H_\lambda+\kappa|\psi|^2\right)\psi\,,
   &&\psi(0)=\psi_0 \label{eq:opt1}\\
   i\dot p&=\left(H_\lambda+2\kappa|\psi|^2\right)p+
   \kappa\psi^2 p^*\,, &&ip(T)=\psi(T)-\psi_d \label{eq:opt2}\\
   \gamma\ddot\lambda&=-\mbox{Re}\Bigl<p\Bigr|
   \frac{\partial H_\lambda}{\partial\lambda}\Bigl|\psi\Bigr>\,,
   &&\lambda(0)=\lambda_0\,,\,\,\lambda(T)=\lambda_1\,. \label{eq:opt3}
\end{align}
Equation.~\eqref{eq:opt1} is the initial value problem of the Gross-Pitaevskii equation, whereas Eq.~\eqref{eq:opt2} is a terminal value problem for the adjoint variable $p$.  Finally, Eq.~\eqref{eq:opt3}  determines the optimal control and is a boundary value problem where both the initial and
terminal value are fixed.

\begin{figure}
\begin{pdffigure}
\centerline{\includegraphics[width=0.6\columnwidth]{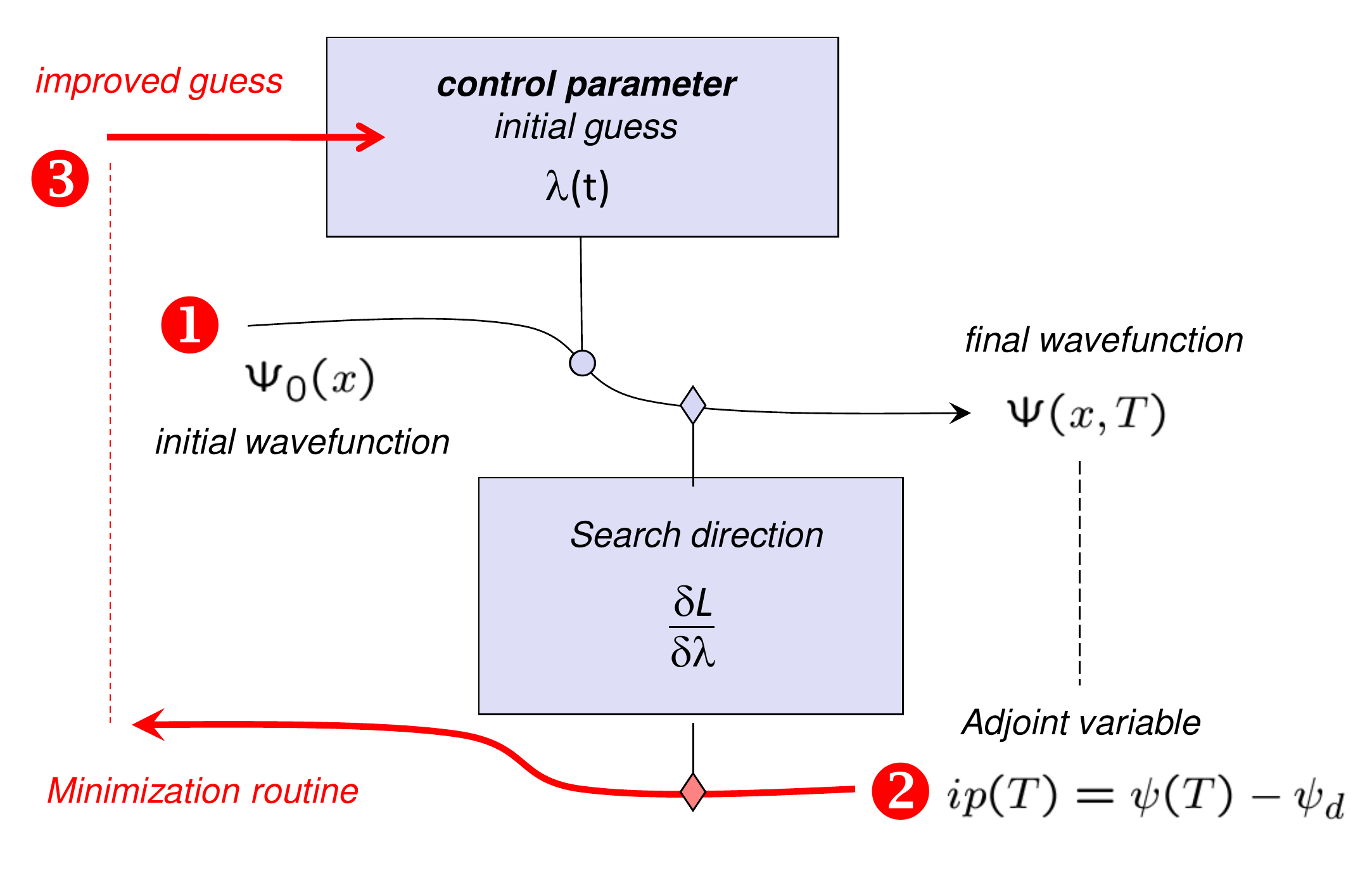}}
\end{pdffigure}
\caption{Schematics for the OCT loop using the Gross-Pitaevskii equation.  For some initial guess for the control parameter $\lambda(t)$, we (1) solve the Gross-Pitaevskii equation forwards in time.  The terminal wavefunction $\psi(T)$ determines the value of the cost function, which becomes minimized within OCT, and allows to (2) compute the terminal condition for the adjoint variable $p(T)$, which is integrated backwards in time.  From the knowledge of $\psi(t)$ and $p(t)$ we can compute the search direction $\delta L/\delta\lambda$ for the control parameter, and (3) come up with an improved guess for the control parameter.  The loop is iterated until a given maximal number of iterations is achieved or the minimum of the cost function is reached. }\label{fig:octschematics}
\end{figure}

In most cases of interest it is impossible to guess an optimal control such that all equations (\ref{eq:opt1}--\ref{eq:opt3}) are fulfilled simultaneously.  However, the above set of equations can be also used for an iterative procedure.  Suppose that we start with some reasonable guess for $\lambda(t)$.  We can then solve Eq.~\eqref{eq:opt1} forwards in time, determine the terminal condition $p(T)$, and solve Eq.~\eqref{eq:opt2} backwards in time.  For a non-optimal control Eq.~\eqref{eq:opt3} is no longer fulfilled.  However,
\begin{equation}\label{eq:opt4}
   \frac{\delta L}{\delta\lambda}=-\gamma\ddot\lambda-
   \Bigl<p\Bigr|
   \frac{\partial H_\lambda}{\partial\lambda}\Bigl|\psi\Bigr> 
\end{equation}
then provides us with a search direction for an optimized control.  Upon searching in the direction $\delta L/\delta\lambda$ we obtain a better control, that brings the terminal $\Psi(T)$ closer to the desired $\psi_d$, and the optimization can be repeated until we reach the minimum of the \textit{optimal control}.  Figure~\ref{fig:octschematics} schematically depicts the OCT loop, which continues until a given maximal number of iterations is achieved or the minimum of the cost function is reached.  Quite generally, for the optimization we either use a nonlinear conjugate gradient method or a quasi-Newton BFGS scheme.

The cost function of Eq.~\eqref{eq:costtrap} for wavefunction trapping is overly restrictive as it requires that the final wavefunction matches the desired one up to the global phase, which is of no relevance.  A better choice $J(\psi)=\frac 12\left[1-|\left<\psi_d|\psi(T)\right>|^2\right]$ for the cost function has been formulated in Ref.~\cite{hohenester.pra:07} and is also discussed in the help pages.  In general, the choice of the cost function is determined by the problem under study.  Cost function and terminal condition for the adjoint variable are connected through
\begin{equation}
  p(T)=-2i\frac{\delta J(\psi,\lambda)}{\delta\psi^*(T)}\,.
\end{equation}
Within the OCTBEC toolbox the terminal condition must be computed by the user through functional derivatives.  However, even if one feels uneasy with functional derivatives this step is usually not overly difficult.  In Sec.~\ref{sec:optgp} we will present a test scheme for figuring out whether the functional derivative and the implementation have been performed properly.

\subsubsection{$L^2$ versus $H^1$ norm}

In some cases, there is a problem related to the numerical implementation of Eq.~\eqref{eq:opt3}.  It is a boundary value problem, where the conditions $\lambda(0)=\lambda_0$ and $\lambda(T)=\lambda_1$ hold.  For the initial guess one often ends up in a state that is quite far away from the desired solution. In turn, the terminal value for $p(T)$ is large and so is the gradient $\delta L/\delta\lambda$ of Eq.~\eqref{eq:opt4}.  This has the consequence that the largest variation of the control field is initially at the terminal time $T$, while from a control perspective it would often be better to change the control at early times first. Only with increasing number of iterations in the minimization procedure the control becomes modified within the whole time interval.  This can lead to a tedious and time-consuming optimization process.  A more convenient approach was formulated by von Winckel and Borz\`\i~\cite{vonwinckel:08}.  The main idea is to use a different norm in the integrals for the cost as well as in the Lagrange function.  In the cost functions the penalization for the control field $(\gamma/2)\,\bigl (\dot\lambda, \dot\lambda\bigr)_{L^2}$ can be reformulated as $(\gamma/2)\,\bigl (\lambda, \lambda\bigr)_{H^1}$, where the definition of the $H^1$ inner product is $(u,v)_{H^1}=(\dot u,\dot v)_{L^2}$.  It is important to realize that this different norm does neither affect the value of the cost function nor the principal or sensitivity Eqs.~(\ref{eq:opt1},\ref{eq:opt2}).  However, it does affect the equation for the control field, which now satisfies a Poisson equation
\begin{equation}\label{eq:opt5}
   -\frac{d^2}{dt^2}\left[\frac{\delta L}{\delta\lambda}\right]=
   -\gamma\ddot\lambda-\mbox{Re}
   \bigl<p\bigr|\frac{\partial H_\lambda}{\partial\lambda}
   \bigl|\psi\bigr> \,.
\end{equation}
The advantage of this equation is that changes due to large values of the second term on the right-hand side are distributed, through the solution of the Poisson equation, over the whole time interval.  Most of our optimizations show that the $H^1$ optimization is faster and more robust than the $L^2$ optimization, the resulting control fields are significantly smoother, and the global structure of $\lambda(t)$ is optimized.

\subsubsection{General OCT problem}

A general OCT problem is governed by a set of differential equations, which we might write as
\begin{equation}
  \dot \psi(t)=f(\psi(t),\lambda(t))\,,\quad \psi(0)=\psi_0\,,
\end{equation}
where $\psi(t)$ are the dynamic variables, such as wavefunctions or orbitals, and $f(\psi(t),\lambda(t))$ is a general functional that determines the time evolution of $\psi(t)$.  The cost function $J(\psi,\lambda)$ can then depend on the terminal value $\psi(T)$ as well as on the intermediate values $\psi(t)$.  We assume that one can separate the cost function into a terminal and intermediate part
\begin{equation}
  J(\psi,\lambda)=J_{\rm final}(\psi(T),\lambda_1)+
  J_{\rm inter}(\psi(t),\lambda(t))+\frac\gamma 2\int_0^T \left(\dot\lambda(t)\right)^2\,dt\,. 
\end{equation}
In practically all our OCT calculations used in the past and given below we set the intermediate cost to zero.  However, for completeness and for future use we have implemented the general scheme, consisting of: 

\medskip
\begin{tabular}{cl}
-- &the final cost function $J_{\rm final}(\psi(T),\lambda_1)$,\\
-- &the terminal conditions $p(T)=-2i[\delta J_{\rm final}/\delta\psi^*(T))]$ 
for the adjoint variables,\\
-- &the intermediate cost function $J_{\rm inter}(\psi,\lambda)$,\\
-- &the functional derivative $f(t)=2[\delta J_{\rm inter}/\delta\psi^*(t))]$,\\
-- &the functional derivative $[\delta J_{\rm inter}/\delta \lambda(t))]$.\\
\end{tabular}
\medskip

\begin{table}
\caption{Classes and class methods for the implementation of optimal control within the framework of the OCTBEC toolbox.  The \texttt{optimality} object solves the optimality Eqs.~(\ref{eq:opt1}--\ref{eq:opt5}), and returns either the value of the cost function or a new search direction.  The \texttt{optimize} object is a generic optimization class using nonlinear conjugate gradient or quasi-Newton BFGS techniques, assuming $L^2$ or $H^1$ norm for the inner products.  Detailed information about the classes can be obtained through \texttt{doc @classname}. }\label{table:oct}
{\small
\begin{tabularx}{\columnwidth}{llX}
\hline\hline
Object & Matlab implementation & Description \\
\hline
Model & \texttt{deriv(psi,lambda,op)} & Derivative function for forward Eq.~\eqref{eq:opt1}\\
& \texttt{crank(psi,lambda,dt,op)} & Crank-Nicolson step for forward Eq.~\eqref{eq:opt1}\\
& \texttt{optderiv(psi,p,lambda,funiter,op)} & Derivative function for adjoint Eq.~\eqref{eq:opt2}\\
& \texttt{optcrank(psi,p,lambda,funiter,dt,op)} & Crank-Nicolson step for adjoint Eq.~\eqref{eq:opt2}\\
& & \\
\texttt{control} & \texttt{control(t,lambda,penalty,flag)} & Control function initialization\\
& \texttt{cost(lambda)} & Cost $(\gamma/2)(\lambda,\lambda)_{H^1}$ for control parameter\\
& \texttt{dlambda=deriv(lambda,f)} & Solve Eq.~\eqref{eq:opt4} or Eq.~\eqref{eq:opt5} \\
& \texttt{ip=inner(u,v )} & Inner product with $L^2$ or $H^1$ norm \\
& \texttt{norm(lambda)} & Norm of control parameter\\
& \texttt{+,-,*,/} & Basic arithmetic operations for one or two control parameters\\
& & \\
\texttt{controltype} & \texttt{lam=lambda(t)} & Return value of control parameter object\\
& \texttt{lam} or \texttt{lam(:)} & Control values at time \texttt{t}\\
& \texttt{lam.t} & Time argument \\
& \texttt{lam.lambda} & Reference to control parameter object\\
& & \\
\texttt{costfunction} & \texttt{costfunction(valfin,final,valint,inter,deriv)} & Initialization of \texttt{costfunction} object \\
&  \texttt{deriv(cost,psi,p,lambda)} & Derivative of cost function wrt $\lambda$\\
& \texttt{p=final(cost,psi,lambda)} & Terminal condition for adjoint variable \\
& \texttt{inter(cost,psi,lambda)} & Inhomgeneity for adjoint equation due to intermediate cost \\
& \texttt{valfin(cost,psi,lambda)} &  Cost value at final time \\
& \texttt{valint(cost,tout,psiout,lambda)} & Cost value at intermediate times \\
& \texttt{*,+,/} & Basic arithmetic operations for cost functions\\
& & \\
\texttt{optimality} & \texttt{optimality(psi0,tout,lambda,costfun,op)} & Initialization of optimality system \\
& \texttt{[val,fin]=cost(opt)} & Cost value and terminal condition for adjoint variable\\
& \texttt{opt=solve(opt)} & Solve forward Eq.~\eqref{eq:opt1}\\
& \texttt{dlambda=isolve(opt)} & Compute search direction for control parameter\\
& & \\
\texttt{optimize} & \texttt{optimize(opt,op)} & Initialize OCT object \\
& \texttt{[oct,lambda,psiout]=improve(obj,niter)} & Perform OCT loop \\
& \texttt{info(oct)} & Print statistics for OCT loop \\
\hline
\hline
\end{tabularx}}
\end{table}

\noindent Table~\ref{table:oct} summarizes the basic ingredients of our OCT implementation.

\subsection{Cost function}

We next discuss how to implement the cost function within the OCTBEC toolbox.  To this end, we have to extend the initialization of the control function (see Sec.~\ref{sec:control}) according to 
\begin{code}
gamma = 1e-2;                                   %  control penalization
lambda = control( ttab, lamtab, gamma, 'H1' );  %  initialize control parameter
\end{code}
Here \verb|ttab| and \verb|lamtab| are the tabulated control parameters for the initial guess, \verb|gamma| is the control penalization, and the last parameter is either \verb|'H2'| or \verb|'L2'| and determines the norm for the inner products.  Once this initialization is performed, the \verb|control| object is ready for use within OCT simulations.  In addition to the cost for the control parameter, we also have to specify the cost e.g. for wavefunction trapping.  Suppose that \verb|psid| is Gross-Pitaevskii object for the desired state.  We then define a \verb!costfunction! object
\begin{code}
%  cost function for state trapping
valfin = @( psi, lambda ) ( 0.5 * ( 1 - abs( grid.inner( psid.val, psi.val ) ) ^ 2 ) );
%  terminal condition for state trapping
final = @( psi, lambda ) ( 1i * grid.inner( psid.val, psi.val ) * psid ); 
%  cost function object
cost = costfunction( valfin, final );
\end{code}
\verb|valfin| gives the cost function for the state trapping, and \verb|final| returns a \verb|grosspitaevskii| object that holds the terminal conditions for the adjoint variable.  In case of an additional intermediate cost, one additionally has to provide the functions \verb|valint|, \verb|inter|, and \verb|deriv| for the intermediate cost, as well as the derivatives of the intermediate cost with respect to the dynamic variables and the control parameters.  The complete cost function definition is of the form
\begin{code}
%  initialize cost function with terminal and intermediate cost
cost = costfunction( valfin, final, valint, inter, deriv );
\end{code}
\verb|costfunction| objects can be multiplied with constant factors, and can be added together.  In the latter case the OCT optimization will come up with control parameters that bridge between the different control objectives.

\subsection{Optimality system}

The OCTBEC toolbox uses an \verb!optimality! object to compute the forward and backward equations, and an \verb!optimize! object to perform the OCT loop.  For the model systems considered in the toolbox we set up the optimality system, e.g., with 
\begin{code}
op1 = struct( 'nsub', 2, 'nout', 0, 'stepfun', 'crank' );  %  options for optimality system
opt = optimality( psi0, tout, lambda, cost, op1, 1 );      %  set up optimality system
\end{code}
The option arguments are for the solution of the dynamic equations and are given in Sec.~\ref{sec:odesolver}.  Additional \verb|op| arguments will be passed to the derivative functions, e.g. to control the tolerance for Newton iterations.  In the initialization of the \verb|optimality| object we provide the initial value for the dynamic variable, the time arguments, the control parameter, the cost function, and the options for the ODE solution.  The last argument will give an intermediate plotting of the control parameters during the OCT loop, and can be omitted or replaced by another function or function handle.  In most cases the \verb|optimality| object is only used internally (see table~\ref{table:oct} for details).

Finally, the OCT loop is performed with an \verb!optimize! object.  It is initialized through
\begin{code}
op2 = struct( 'mode', 'BFGS', 'tol', 1e-4 );  %  options for minimization
oct = optimize( opt, op2 );                   %  set up optimal control object
\end{code}
The following options can be passed to the \verb|optimize| object

\medskip
\begin{tabular}{ll}
\texttt{mode} & \verb!'grad'! for nonlinear conjugate gradient and \verb!'BFGS'! for quasi-Newton optimization, \\
\texttt{tol} & tolerance for the termination of the OCT loop,\\
\texttt{bounds} & \verb![lmin,lmax]! for lower and upper bounds for the control parameter.\\
\end{tabular}
\medskip

\noindent In general, \verb|grad| performs significantly faster than \verb|BFGS| but usually gives slightly worse control fields.  We recommend to use \verb|BFGS| whenever possible.  Bounds can be provided for BFGS simulations in order to avoid too large control fields, for which the ODE integration may fail.  In general we recommend to use not too tight bounds.  If a third value \verb|[lmin,lmax,dlmax]| is provided, the control parameters are also restricted to $|\dot\lambda|<$\texttt{dlmax}.  Without bounds the linesearch algorithm \verb!fminunc! is used, and \verb!fmincon! otherwise.  Additional options to the \verb|optimize| object are passed to these functions.

The OCT loop is then performed through
\begin{code}
[ oct, lambda, psiout ] = oct.improve( 10 );  %  perform OCT iterations
\end{code}
In this case either 10 OCT iterations are performed, or the OCT loop terminates when $\|\delta L/\delta\lambda\|$ becomes smaller than \verb|tol| times its initial value.  Upon termination of the OCT loop, we obtain 

\medskip
\begin{tabular}{ll}
\texttt{oct} & updated OCT object, \\
\texttt{lambda} & optimized control parameter,\\
\texttt{psiout} & history for dynamic variables as computed by the ODE solver.\\
\end{tabular}
\medskip

\noindent If one wants to continue with the optimization, one can call again \verb|oct.improve(niter)| to further improve the optimal control.  Finally, some information about the optimization is availabe through
\begin{code}
>> oct.info;  %  give information about OCT loop

Optimal control loop :

Number of iterations: 10
Search algorithm:     BFGS

Total elapsed time:              43.426825
Mean time for  forward solution: 0.240128
Mean time for backward solution: 1.179120

Time percentage for  forward solution: 68.566
Time percentage for backward solution: 27.152
Time percentage for rest:              4.282

Number of  forward solutions:     124
Number of backward solutions:     10
\end{code}

\section{Optimization of model systems}\label{sec:optmodel}

\subsection{Optimization of the Gross-Pitaevskii equation}\label{sec:optgp}

The demo program \verb!demogpoct1.m! provides an example for the optimization of a splitting process within the framework of optimal control theory.  The basic steps are:

\medskip
\begin{tabular}{cl}
1. & set up computational grid and Hamiltonian, and compute initial
wavefunction,\\
2. & define time interval and initial guess for control parameter,\\
3. & define cost properties for control (penalization and $L^2$ or $H^1$ norm) and
cost function,\\
4. & set up options for solution of Gross-Pitaevskii equation,\\
5. & submit problem to OCT minimization.\\
\end{tabular}
\medskip

\noindent Steps 1,2 and 4 are similar to the solution of the Gross-Pitaevskii equation, and step 5 in general involves just a few program lines.  The definition of the cost function and its derivative of step 3 provide the main challenge of OCT simulations.  Typically the implementation of step 3 is relatively simple, but has to be done with care.  In our demo program we first set up the computational grid and Hamiltonian, and define the nonlinearity.
\begin{code}
units;                         %  internal units
grid = grid1d( - 3, 3, 101 );  %  simulation grid
v = @( lambda ) ( lesanovsky1d( grid.x, lambda ) );  %  confinement potential
ham = @( lambda ) ( - 0.5 * grid.lap4 / ( 87 * mass ) + spdiag( v( lambda ) ) );
                               %  single particle Hamiltonian
kappa = pi;                    %  nonlinearity
\end{code}
The above steps have been discussed in length in the Sec.~\ref{sec:preliminaries} and \ref{sec:gp}.  We next define the time interval of the optimization and define an initial guess for the control parameter.
\begin{code}
tmax = 1.2;                       %  maximal time of simulation
tout = linspace( 0, tmax, 100 );  %  output times

gamma = 1e-2;                         %  penalization of control field
lambda = sqrt( tout / max( tout ) );  %  control parameter
lambda = control( tout, lambda, gamma, 'H1' );
\end{code}
In the last line we set up a \verb!control! object using the initial guess for the control parameter.  The last two parameters are \verb|gamma| for the penalization of the control parameter, and \verb|'L2'| or \verb|'H1'| for the norm used in the OCT simulations (we recommend to always use \verb|'H1'|).  Next, we compute the Gross-Pitaevskii groundstate and the desired OCT state, within which the system should end up after the splitting process.  Here we define the desired state as the groundstate of the split trap.  With the ground and desired state we can now set up a \verb!costfunction! object.
\begin{code}
%  groundstate wavefunction
psi0 = groundstate( grosspitaevskii( grid, ham, kappa ), lambda.first );
%  desired wavefunction at terminal time
psid = groundstate( grosspitaevskii( grid, ham, kappa ), lambda.last, 'mix', 1e-2 );

%  cost function for state trapping
value = @( psi, lambda ) ( 0.5 * ( 1 - abs( grid.inner( psid.val, psi.val ) ) ^ 2 ) );
%  terminal condition for adjoint variable
final = @( psi, lambda ) ( 1i * grid.inner( psid.val, psi.val ) * psid ); 
%  initialize cost function object
cost = costfunction( value, final );
\end{code}
We next define the options for the ODE solver (see Sec.~\ref{sec:odesolver}) and define the optimality system.
\begin{code}
%  options for ODE solver
op1 = struct( 'nsub', 2, 'nout', 0, 'stepfun', 'crank' );
%  set up optimality system
opt = optimality( psi0, tout, lambda, cost, op1, 1 );
\end{code}
In the help pages we show how the \verb!optimality! object can be used to compute the search direction for the optimal control.  Finally, the optimal control loop is performed with the following commands.
\begin{code}
op2 = struct( 'mode', 'BFGS', 'tol', 1e-4 );  %  options for OCT minimization
oct = optimize( opt, op2 );                   %  set up optimal control object
[ oct, lambda, psiout ] = oct.improve( 10 );  %  perform OCT loop
\end{code}
The first command defines the options for the OCT minimization.  We use a quasi-Newton scheme and use a tolerance of \verb|tol=1e-4| for the termination of the OCT loop.  The call \verb|oct.improve(10)| performs the OCT optimization, which ends after 10 iterations or when the tolerance is reached.  Further iterations can be performed by either increasing the number of iterations (e.g. from 10 to 100) or by calling again the \verb|oct.improve| function.  As the BFGS algorithm uses the previous function calls in order to estimate the Hessian of the control space, the two approaches will give slightly different results.  If possible, we recommend to increase the number of iterations.  Results of the OCT simulation have already been shown in the introductory Sec.~\ref{sec:examples} and in Fig.~\ref{fig:examples}.

The functional derivative of the cost function must be implemented properly.  Otherwise the OCT algorithm will not terminate properly, usually because the cost function does not decrease along the search direction, or run forever.  It is usually a good idea to make consistency checks, as shown in the demo program \verb!demogpoct1test.m!.  The idea is as follows.  Suppose that $\lambda(t)$ is a control field and $u(t)$ some smooth function with the boundary conditions $u(0)=u(T)=0$.  Then, for a small parameter $\eta$ the following relation holds
\begin{equation}
  \frac 1\eta\Bigl[J(\psi,\lambda+\eta u)-J(\psi,\lambda)\Bigr]
  =\left< u(t),\frac{\delta L(\psi,p,\lambda)}{\delta\lambda(t)}
  \right>_{L^2\,\,{\rm or}\,\,H^1}+O(\eta)\,, 
\end{equation}
which can be used to test the OCT implementation.  
\begin{code}
%  small quantity
eta = 1e-6;
%  variation of control parameter
dlambda = control( tout, sin( tout / max( tout ) * 6 * pi ), 1e-10, 'L2' );

%  options for ODE solver
op = struct( 'nsub', 2, 'nout', 0, 'stepfun', 'crank' );
%  set up optimality systems
opt1 = optimality( psi0, tout, lambda,                 cost, op );
opt2 = optimality( psi0, tout, lambda + eta * dlambda, cost, op );
%  solve Gross-Pitaevskii equation and compute cost function
opt1 = solve( opt1 );  j1 = opt1.cost();
opt2 = solve( opt2 );  j2 = opt2.cost();

deriv = isolve( opt1 );        %  OCT derivative
dj = inner( dlambda, deriv );  %  cost difference from OCT derivative

%  final output
fprintf( 'Direct:   %9.5f\n', ( j2 - j1 ) / eta );
fprintf( 'OCT:      %9.5f\n', dj );

  Direct:    -0.35799
  OCT:       -0.35832
\end{code}
In general the two results will somewhat differ because of numerical rounding errors.  However, significant differences between these two computation approaches usually indicate improper OCT implementations which must be corrected.  We recommend to always employ this testing approach before running the OCT loop. 

\subsection{Optimization of the multi-configurational Hartree method for bosons (MCTDHB)}\label{sec:optmctdhb}

Optimization of control fields within the MCTDHB framework is very similar to the optimization within the Gross-Pitaevskii framework, although the underlying equations are more difficult to solve.  In \ref{sec:eqsoptmctdhb} we provide some details about the working equations.  The choice of the cost function within MCTDHB is less obvious, in \texttt{demomctdhb1.m} we provide an example for orbital trapping and in \texttt{demomctdhb2.m} for energy minimization of the terminal states.  See also Refs.~\cite{grond.pra:09,grond.pra:09b,hohenester.fdp:09,grond.njp:10,grond:11,grond:11b} for further examples.  The programs perform significantly slower than for Gross-Pitaevskii OCT simulations, with typical runtimes listed in table~\ref{table:examples}.

\subsection{Optimization of generic few-mode models}\label{sec:optfewmode}

\begin{figure}
\centerline{\includegraphics[width=0.35\columnwidth]{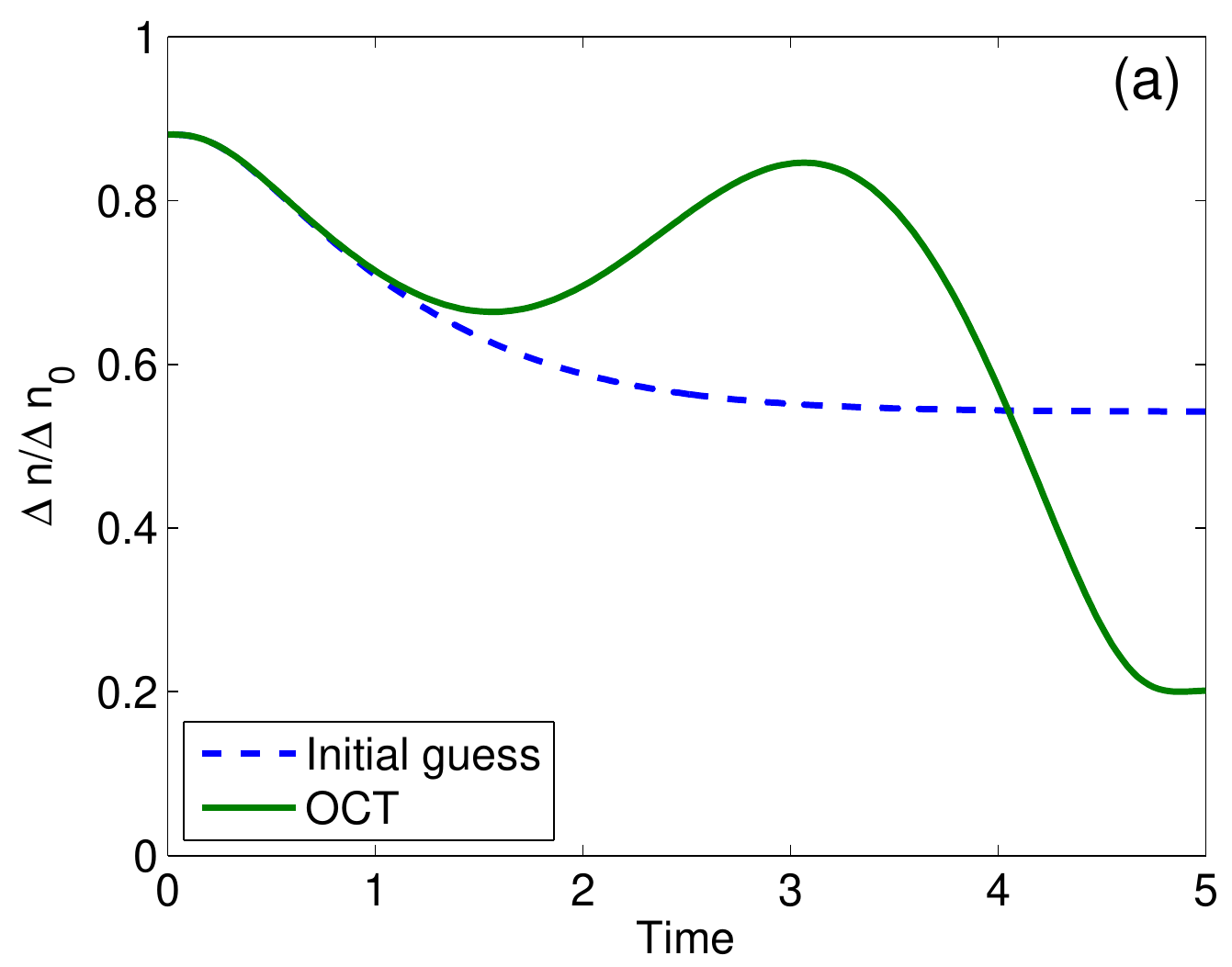}\quad
            \includegraphics[width=0.35\columnwidth]{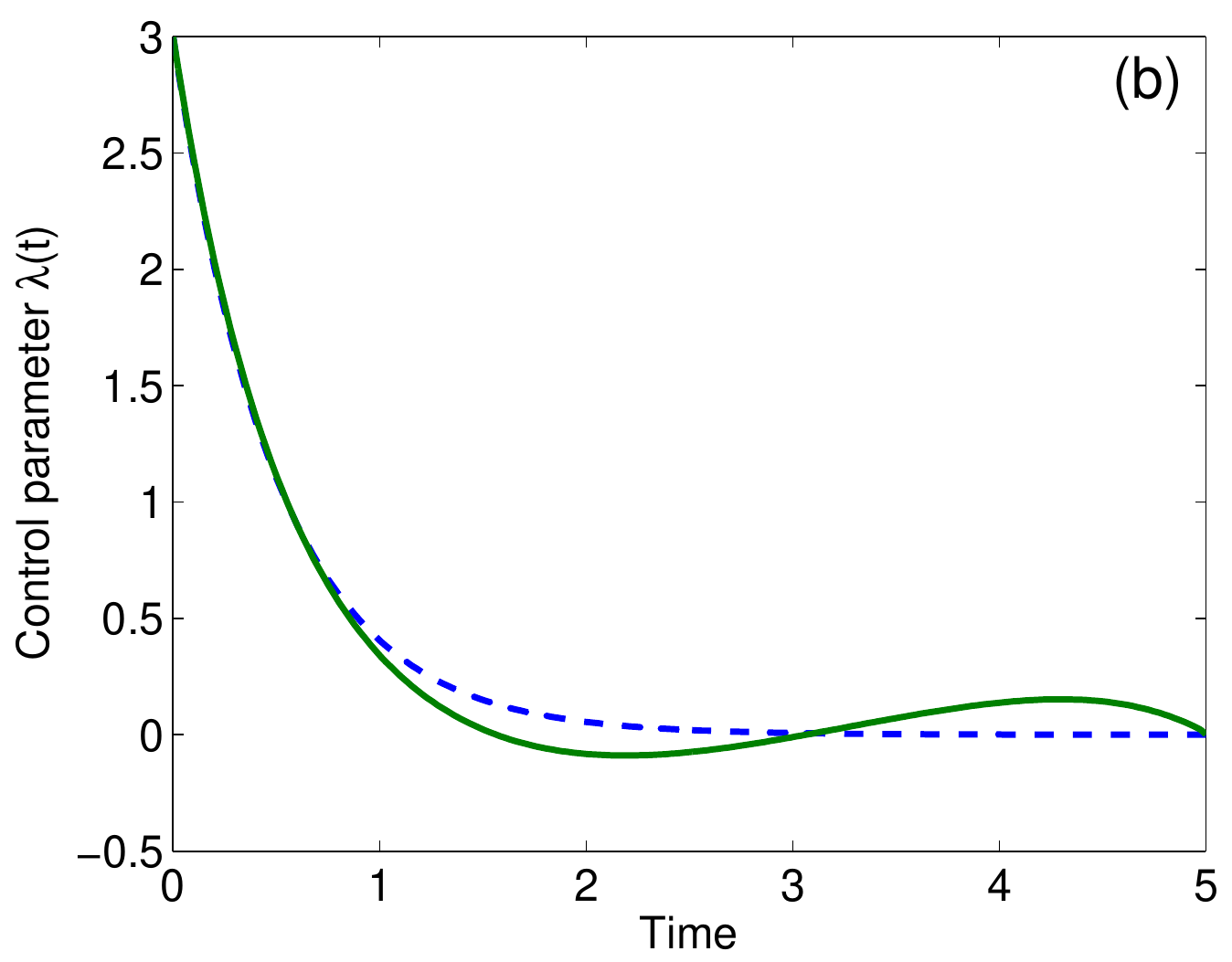}}
\caption{Optimization of number squeezing for two-mode model and Hamiltonian of Eq.~\eqref{eq:fewmode}.  (a) Density fluctuation in units of $\sqrt n/2$ for initial guess (dashed line) and optimized control parameter (solid line), as computed with demo file \texttt{demofewmodepairoct.m}.  Panel (b) reports the corresponding control parameters.}\label{fig:fewmodeoct}
\end{figure}

We finally briefly comment on OCT optimization for the atom number part of the wavefunction.  In Sec.~\ref{sec:fewmode} we have discussed the demo file \texttt{demofewmodepair.m} where within a two-mode model atom number squeezing is achieved by adiabatically turning off the tunneling coupling.  In \texttt{demofewmodepairoct.m} we present an OCT optimization for such number squeezing, as also discussed in Ref.~\cite{grond.pra:09,grond.pra:09b}.  Our goal is to minimize in the terminal state atom-number fluctuations between the left and right well.  The atom number imbalance between the two wells is measured by the pseudospin operator
\begin{equation}
  \hat J_z=\frac 12\left(\hat a_1^\dagger \hat a_1-\hat a_2^\dagger \hat a_2\right)\,. 
\end{equation}
The control objective is to minimize atom number fluctuations in the terminal state, once tunneling has been switched off.  To this end, we introduce the cost function
\begin{equation}
 J(C)=\left<\hat J_z^2\right>-\Bigl(\left<\hat J_z\right>\Bigr)^2=
        \langle C|\hat J_z^2|C\rangle \,. 
\end{equation}
Here $C$ is the atom-number wavefunction, and to arrive at the last expression we have used $\left<\hat J_z\right>=0$.  Taking the functional derivative of the cost function with respect to $C^\dagger$, we arrive for the adjoint variable $D$ at the terminal condition $D(T)=-2i \hat J_z^2\, C(T)$.  Within our program, we implement the \texttt{costfunction} object in the general way
\begin{code}
jz = spin.op( 0.5 * [ 1, 0; 0, - 1 ] );  %  pseudospin operator

%  cost function for squeezing (use REAL to avoid rounding errors)
value = @( psi, lambda ) real( psi.num' * ( jz ^ 2 ) * psi.num - ( psi.num' * jz * psi.num ) ^ 2 );
%  derivative of cost function
final = @( psi, lambda ) ( - 2i * ( jz ^ 2 - 2 * ( psi.num' * jz * psi.num ) ) * psi ); 
%  cost function object
cost = costfunction( value, final ) / ( n / 4 );
\end{code}
Division with $n/4$ ensures that we measure the (square of the) fluctuations with respect to the fluctuations of a binomial state.  We next set up the optimality system and the optimal control object.
\begin{code}
op = struct( 'nsub', 2, 'nout', 0, 'stepfun', 'crank' );  %  options for ODE solver
opt = optimality( psi0, tout, lambda, cost, op, 1 );      %  set up optimality system

op = struct( 'mode', 'BFGS', 'tol', 1e-6 );   %  options for OCT minimization
oct = optimize( opt, op );                    %  set up optimal control object
[ oct, lambda, psiout ] = oct.improve( 10 );  %  perform OCT iterations

it=  1   f=3.367686e-001   ||g||=3.623906e+001   sig=0.012
it=  2   f=8.584814e-002   ||g||=1.041520e-001   sig=0.661
it=  3   f=8.532199e-002   ||g||=3.253682e-004   sig=0.010
it=  4   f=8.454468e-002   ||g||=1.456896e-001   sig=5.642
it=  5   f=8.453521e-002   ||g||=1.776887e-001   sig=0.001
it=  5   f=8.453521e-002   ||g||=1.776887e-001
\end{code}
\noindent Figure~\ref{fig:fewmodeoct} shows the time evolution of the atom-number fluctuations measured in units of the fluctuations of a binomial state.  As regarding the interpretation of the control strategy we refer to Ref.~\cite{grond.pra:09b}.  It is also instructive to visualize the
state evolution on the Bloch sphere, as can be done by uncommenting the last lines in \texttt{demogpoctfewmodepair.m}.

\section*{Acknowledgment}

I am grateful to J\"org Schmiedmayer, Julian Grond, Georg J\"ager, and Robert Sch\"utky for most helpful discussions.  Alfio Borz\`\i\ and Greg von Winckel are cordially acknowledged for their contributions to the OCT implementation.  This work has been supported in part by the Austrian science fund FWF under project P24248.

\appendix

\section{Crank-Nicolson scheme}\label{sec:crank}

In this appendix we provide some details for the solution of Schr\"odinger-type equations with the Crank-Nicolson scheme.  We will derive the pertinent expressions needed for the solution of the Gross-Pitaevskii equation, the MCTDHB equations, and the fewmode model.

\subsection{Schr\"odinger equation}

For the sake of completeness, we briefly review the solution of the Schr\"odin\-ger equation with the Crank-Nicolson scheme.  Let $\psi^n$ denote the wavefunction at time $t_n$.  The time derivative is approximated by the time-discretized version ($\hbar=1$)
\begin{equation}\label{eq:schroedingercrank}
  \frac i{\Delta t}\left(\psi^{n+1}-\psi^n\right)=H
  \frac{\psi^n+\psi^{n+1}}2\,,
\end{equation}
where $H$ is a sparse matrix consisting of the kinetic and potential term evaluated at the mid-time $\frac{t_n+t_{n+1}}2$.  Note that for simplicity we have not indicated the space dependence of the wavefunction and the Hamiltonian matrix.  In the solution of Eq.~\eqref{eq:schroedingercrank} we know the wavefunction $\psi^n$ at the initial time and are seeking for the wavefunction $\psi^{n+1}$ at the end of the time step.  Solving Eq.~\eqref{eq:schroedingercrank} for the unknown then gives the final expression
\begin{equation}\label{eq:schroedingercranksol}
  \psi^{n+1}=\left(\openone+\frac i{2\Delta t}H\right)^{-1}
             \left(\openone-\frac i{2\Delta t}H\right)\psi^n\,.
\end{equation}
The solution of Eq.~\eqref{eq:schroedingercranksol} is simplified by the sparseness of the Hamiltonian matrix which allows for a fast inversion through a LU decomposition \cite{press:02}.

\subsection{Gross-Pitaevskii}

Submitting the Gross-Pitaevskii equation \eqref{eq:gp} to the Crank-Nicolson discretization, we obtain
\begin{equation}\label{eq:gpcrank}
  \frac i{\Delta t}\left(\psi^{n+1}-\psi^n\right)=H
  \frac{\psi^n+\psi^{n+1}}2+\kappa \left|\frac{\psi^n+\psi^{n+1}}2\right|^2
  \frac{\psi^n+\psi^{n+1}}2\,.
\end{equation}
Due to the atom-atom interaction the equation becomes nonlinear and cannot be solved straightforwardly.  We here follow the approach presented in Ref.~\cite{grond.pra:09b} which solves Eq.~\eqref{eq:gpcrank} iteratively through a Newton iteration scheme.  In absence of the nonlinearity we first use Eq.~\eqref{eq:schroedingercranksol} to obtain an approximate solution $\psi_0^{n+1}$.  To obtain the true solution $\psi^{n+1}\equiv\psi_0^{n+1}+\delta\psi$, we next linearize Eq.~\eqref{eq:gpcrank} with respect to $\delta\psi$,
\begin{equation}
  \frac i{\Delta t}\left(\psi_-+\delta\psi\right)=H\left(\psi_++\frac{\delta\psi}2\right)+
  \kappa\left|\psi_+\right|^2\left(\psi_++\delta\psi\right)
  +\kappa\psi_+^2\frac{\delta\psi^*}2\,,
\end{equation}
where we have introduced the abbreviations $\psi_\pm=\psi_0^{n+1}\pm\psi^n$.  Separating the expressions with $\delta\psi$ from the rest, we get
\begin{equation}\label{eq:gpcranksol}
  \left(\openone+i\frac{\Delta t}2\left[H+2\kappa\left|\psi_+\right|^2\right]\right)\delta\psi+
  i\frac{\Delta t}2\kappa \psi_+^2\delta\psi^*=
  -\psi_--i\Delta t\left(H+\kappa\left|\psi_+\right|^2\right)\psi_+\,.
\end{equation}
This equation is of the form $Ax+Bx^*=b$, where $A$ and $B$ are matrices, $x$ is the solution vector, and $b$ an inhomogeneity.  Together with the complex conjugate of the equation we obtain the matrix equation
\begin{equation}
  \begin{pmatrix} A & B \\ A^* & B^* \\ \end{pmatrix}
  \begin{pmatrix}x \\ x^* \end{pmatrix}=\begin{pmatrix}b \\ b^* \end{pmatrix}\,,
\end{equation}
which can be solved by inversion and keeping only the $x$ part of the solution vector.  Due to the sparseness of the matrices $A$ and $B$ the inversion is again fast and efficient.

In our computational approach, we solve Eq.~\eqref{eq:gpcranksol} iteratively, by replacing in a second step the initial guess with $\psi_0^{n+1}+\delta\psi$ and subsequent solution of Eq.~\eqref{eq:gpcranksol}, and repeat this procedure until convergence.  With a tolerance \verb|tol=1e-6| for the termination of the Newton iteration loop, we typically need two to three iterations.

\subsection{Schr\"odinger-type equation with inner products}\label{sec:atomcrank}

Let us consider the Schr\"odinger equation
\begin{equation}\label{eq:atomcrank}
  i\frac{\partial\psi}{\partial t}=\bigl(H-\langle\psi|H|\psi\rangle\bigr)\psi\,,
\end{equation}
which e.g. arises for the atom-number part of the wavefunction.  Submitting the equation to the Crank-Nicolson scheme, we get
\begin{equation}\label{eq:atomcranksol}
  \frac i{\Delta t}\left(\psi_-+\delta\psi\right)=
  \bigl(H-\langle\psi_+|H|\psi_+\rangle\bigr)\left(\psi_++\frac{\delta\psi}2\right)-
  \frac 12\bigl(\langle\psi_+|H|\delta\psi\rangle+
                \langle\delta\psi|H|\psi_+\rangle\bigr)\psi_+\,.
\end{equation}
$\psi_0$ is an approximate solution at the end of the time step, which is e.g. obtained by solving Eq.~\eqref{eq:atomcrank} with the subtraction term $\langle\psi^n|H|\psi^n\rangle$, and we have again introduced the abbreviations $\psi_\pm=\psi_0^{n+1}\pm\psi^n$ as well as $\delta\psi$ for the difference between $\psi_0$ and the true solution.

Eq.~\eqref{eq:atomcranksol} is of the form
\begin{equation}
  Ax+Bx^*+(V_1^Tx)U_1+(V_2^Tx^*)U_2=b\,,
\end{equation}
where $A$ and $B$ are matrices (matrix $B$ has been added for completeness), and $U_k$ and $V_k$ are vectors.  Together with the complex conjugate of the above equation we get
\begin{eqnarray}
  \begin{pmatrix} A & B \\ A^* & B^* \\ \end{pmatrix}\begin{pmatrix}x \\ x^* \end{pmatrix}&+&
  (V_1^Tx\phantom{^*})\begin{pmatrix}U_1 \\ 0 \end{pmatrix}+
  (V_1^Tx\phantom{^*})^*\begin{pmatrix}0 \\ U_1^* \end{pmatrix} \\ &+&
  (V_2^Tx^*)\begin{pmatrix}U_2 \\ 0 \end{pmatrix}+
  (V_2^Tx^*)^*\begin{pmatrix}0 \\ U_2^* \end{pmatrix}=
  \begin{pmatrix}b\phantom{^*} \\ b^* \end{pmatrix}\,, \nonumber
\end{eqnarray}
which can be written in the compact form
\begin{equation}
  \tilde A\tilde x+\sum_k \left(\tilde V_k^T\tilde x\right)\tilde U_k=
  \left(\tilde A+\tilde U\tilde V^T\right)\tilde x=\tilde b\,.
\end{equation}
Here $\tilde U=(\tilde U_1,\tilde U_2,\dots)$ and $\tilde V=(\tilde V_1,\tilde V_2,\dots)$ are matrices consisting of the vectors $\tilde U_k$ and $\tilde V_k$.  The matrix $\tilde U\tilde V^T$ is of low rank as the two matrices $\tilde U$ and $\tilde V$ only consists of a few column vectors.  One can thus employ the Sherman--Morrison--Woodbury formula for the inversion~\cite{grond.pra:09b}
\begin{equation}\label{eq:woodbury}
  \left(\tilde A+\tilde U\tilde V^T\right)^{-1}=\tilde A^{-1}-\left[\tilde A^{-1}\tilde U\right]
  \left(\openone+\tilde V^T\left[\tilde A^{-1}\tilde U\right]\right)^{-1} 
  \tilde V^T\tilde A^{-1}\,.
\end{equation}
When applying this matrix to $\tilde x$, we have to solve $\tilde A^{-1}\tilde x$ and $\tilde A^{-1}\tilde U$ which is typically efficient because of the sparseness of the matrix $A$.  The inversion of the matrix on the right-hand side of Eq.~\eqref{eq:woodbury} is also fast because of the small number of $\tilde U_k$ and $\tilde V_k$ vectors involved.  For these reasons, the solution of $(\tilde A+\tilde U\tilde V^T)^{-1}\tilde x$ can be very fast and efficient.  As for the Schr\"odinger equation \eqref{eq:atomcrank}, we then solve Eq.~\eqref{eq:atomcranksol} iteratively with a Newton iteration scheme.  

\subsection{Multi-configurational Hartree method for bosons}

The Sherman--Morrison--Woodbury formula can be also employed in the Crank-Nicolson solution of the MCTDHB equations, which consist of one equation for the atom number part $C$ and another set of equations for the orbitals $\phi_k$.  As for the atom number part, we obtain in correspondence to Eq.~\eqref{eq:atomcranksol} the expression
\begin{equation}\label{eq:mctdhbatomcrank}
  \frac i{\Delta t}\left(C_-+\delta C\right)= 
  \bigl(H-\langle C_+|H|C_+\rangle\bigr)\left(C_++\frac{\delta C}2\right)-
  \frac 12\bigl(\langle C_+|H|\delta C\rangle+
                \langle\delta C|H|C_+\rangle\bigr)C_+\,,
\end{equation}
with the shorthand notations $C_\pm=C_0^{n+1}\pm C^n$ and $\delta C=C^{n+1}-C_0^{n+1}$, where $C_0$ is some initial guess.  Quite generally, the Hamiltonian $H$ depends on the orbitals $\phi_k^{n+1}$ which are not known at the first iteration.  In our computational approach we do not consider this dependence explicitly, but solve Eq.~\eqref{eq:mctdhbatomcrank} in parallel with Eq.~\eqref{eq:mctdhborbitalcrank}, to be derived below.  As the dependence of $H$ on the orbitals is generally rather weak (as well as the dependence of the density matrices on $C$), one can expect that the iterative solution of the coupled atom and orbital equations leads to convergence even without explicit consideration of these dependencies.  Indeed, in all situations considered so far we found convergence after a few iterations.

As for the orbital part, we have to solve
\begin{equation}
  i\dot\phi_i=\mathcal{P}\Bigl[h\phi_i+\kappa\sum_{jkl}r_{ijkl}\phi_j^*\phi_k\phi_l\Bigr]
  \equiv \mathcal{P}f_i\,,\quad r_{ijkl}=\sum_a\rho_{ia}^{-1}\rho_{ajkl}^{(2)}\,.
\end{equation}
Here $\mathcal{P}$ is the projection operator \cite{alon:08}.  Submitting this equation to the Crank-Nicolson scheme, and ignoring the dependence of $r$ on the atom number part $C$, we obtain
\begin{equation}\label{eq:mctdhborbitalcrank}
  i\frac{\phi_i^-+\delta\phi_i}{\Delta t}=f_i^++\delta\!f_i-
  \sum_j\langle\phi_j^+|f_i^++\delta\!f_i\rangle \phi_i^+
  -\frac 12\sum_j\Bigl[\langle\delta\phi_j|f_i^+\rangle \phi_j^++
  \langle\phi_j^+|f_i^+\rangle\delta\phi_j\Bigr]\,,
\end{equation}
with
\begin{displaymath}
  \delta\!f_i=h\frac{\delta\phi_i}2+\kappa\sum_{jkl}r_{ijkl}\left(\phi_j^*\phi_k\delta\phi_l+
  \frac 12\delta\phi_j^*\phi_k\phi_l\right)\,.
\end{displaymath}
Eq.~\eqref{eq:mctdhborbitalcrank} can be solved with the Sherman--Morrison--Woodbury formula of Eq.~\eqref{eq:woodbury}.  In many cases, we neglect in Eq.~\eqref{eq:mctdhborbitalcrank} the last term on the right-hand side, which is due to the projector in the orbital equations.  This can be controlled through the \verb!proj! argument in the options which can bes set to \verb!'on'! or \verb!'off'!.  For two orbitals the Newton iteration converges even without the projector term, and the computation becomes significantly faster.

\section{Optimal control equations for MCTDHB model}\label{sec:eqsoptmctdhb}

Here we derive the somewhat lengthy optimal control equations for the MCTDHB equations.  Let us first consider the atom number Hamiltonian
\begin{equation}
  H=\sum_{ij}\langle i|h|j\rangle\, \hat a_i^\dagger \hat a_j^{\phantom\dagger}+
  \frac\kappa 2\sum_{ijkl}\int \phi_i^*\phi_j^*\phi_k\phi_l\,dx\,
  \hat a_i^\dagger \hat a_j^\dagger \hat a_k^{\phantom\dagger}\hat a_l^{\phantom\dagger}\,.
\end{equation}
Differentiation with respect to the atomic orbitals gives
\begin{equation}
  \frac{\delta H}{\delta\phi_i^*}=\sum_j h\phi_j\,\hat a_i^\dagger \hat a_j^{\phantom\dagger}+
  \kappa\sum_{jkl}\phi_j^*\phi_k\phi_l\,
  \hat a_i^\dagger \hat a_j^\dagger \hat a_k^{\phantom\dagger}\hat a_l^{\phantom\dagger}
  \equiv\sum_\mu\varphi_{i\mu}h_{i\mu}\,,
\end{equation}
where the last expression defines the orbital functions $\varphi_{i\mu}$ and the corresponding atom number operators $h_{i\mu}$.  For a given cost function $J$, the Lagrange function of reads
\begin{eqnarray}
  L&=& \frac 12\int\Bigl[\tilde C^\dagger\left(i\dot C-(H-\bar H)C\right)+
                \left(-i\dot C^\dagger-C^\dagger(H-\bar H)\right)\tilde C\Bigr]\,dt\nonumber\\
  &+&\frac 12\int\sum_i\Bigl[\tilde\phi_i^*\left(i\dot\phi_i-\mathcal{P}f_i\right)+
                      \left(-i\dot\phi_i^*-\mathcal{P}f_i^*\right)\tilde\phi_i\Bigr]\,dxdt
  +J\,,
\end{eqnarray}
where we have introduced the shorthand notation $\bar H=\langle C|H|C\rangle$.  Here $\tilde C$ and $\tilde\phi_i$ are the adjoint variables introduced as constraints in order to fulfill the dynamic MCTDHB equations.   To derive the OCT equations, we have to perform variational derivatives with respect to $C$ and $\phi_i$.  As for the orbital part, we obtain
\begin{eqnarray}
  i\dot{\tilde\phi}_i &=& h\left(\mathcal{P}\tilde\phi_i\right)+
  \kappa\sum_{jkl}r_{ijkl}\left[\left(\mathcal{P}\tilde\phi_j\right)^*\phi_k\phi_l+
  2\phi_j^*\phi_k\left(\mathcal{P}\tilde\phi_l\right)\right]\nonumber\\ &-&
  \sum_j\left(\bigl<\tilde\phi_j\bigr|\phi_i\bigr>f_j-
  \bigl<f_j\bigr|\phi_i\bigr>\tilde\phi_j\right)+
  \sum_\mu\left[\bigl<\tilde C\bigr|h_{i\mu}-\bar h_{i\mu}\bigl|C\bigr>+
  \bigl<C\bigr|h_{i\mu}-\bar h_{i\mu}\bigl|\tilde C\bigr> \right]\phi_{i\mu}\,,
\end{eqnarray}
where $\bar h_{i\mu}=\langle C|h_{i\mu}|C\rangle h_{i\mu}$.  As for the atom-number part of the adjoint equations, we next investigate the derivative of the density matrix term $r=\rho^{-1}\rho^{(2)}$.  From $\rho^{-1}\rho=1$ we find $[\delta\rho^{-1}]\rho+\rho^{-1}\delta\rho=0$, which finally yields $\delta\rho^{-1}=-\rho^{-1}\rho\rho^{-1}$.  Thus, $\delta r=\rho^{-1}[-\delta\rho\,r+\delta\rho^{(2)}]$.  We then find
\begin{equation}
  \frac{\delta}{\delta C^\dagger}\sum_{jkl}r_{ijkl}\phi_j^*\phi_k\phi_l=
  \sum_{jklm}\left[-\left(\rho^{-1}\hat\rho\right)_{im} r_{mjkl}+
  \rho_{im}^{-1}\hat\rho^{(2)}_{mjkl}\right]C\equiv \sum_{\mu}g_{i\mu}r_{i\mu}\,C\,,
\end{equation}
where we have introduced the operators $\hat\rho_{ij}=\hat a_i^\dagger\hat a_j$ and $\hat\rho_{ijkl}^{(2)}=\hat a_i^\dagger\hat a_j^\dagger\hat a_k\hat a_l$.  The adjoint equation for the atom-number part of the wavefunction finally reads
\begin{equation}
  i\dot{\tilde C}=\left(H-\bar H\right)C+2\mbox{Re}\left[\tilde C^\dagger\cdot C\right]
  HC+\sum_{i\mu}\left[\bigl<\tilde\phi_i\bigr|\mathcal{P}g_{i\mu}\bigr>r_{i\mu}+
  \bigl<\mathcal{P}g_{i\mu}\bigr|\tilde\phi_i\bigr>r_{i\mu}^\dagger\right]C\,.
\end{equation}




\end{document}